\documentclass[prb,aps,twocolumn]{revtex4}
\usepackage{amsmath}
\usepackage{graphicx}
\newcommand{\sech}{\ensuremath{\mathrm{sech}}}
\newcommand{\COSH}{\ensuremath{\cosh(R/\xi)}}
\newcommand{\SECH}{\ensuremath{\sech(R/\xi)}}
\newcommand{\SINH}{\ensuremath{\sinh(R/\xi)}}
\begin{document}
\title{Positive current cross-correlations in a highly transparent
  normal-superconducting beam splitter due to synchronized
  Andreev and inverse Andreev reflections}

\date{\today}

\author{Axel Freyn}

\affiliation{Institut NEEL,
  CNRS and Universit\'e Joseph Fourier, BP 166,
  F-38042 Grenoble Cedex 9, France}

\author{Martina Fl\"oser}

\affiliation{Institut NEEL,
  CNRS and Universit\'e Joseph Fourier, BP 166,
  F-38042 Grenoble Cedex 9, France}

\author{R\'egis M\'elin}
\email{Regis.Melin@grenoble.cnrs.fr}
\affiliation{Institut NEEL,
  CNRS and Universit\'e Joseph Fourier, BP 166,
  F-38042 Grenoble Cedex 9, France}

\newcommand {\ARARbar}{\ensuremath {\mathrm {AR}\text{-}\overline{\mathrm
      {AR}}}}
\newcommand {\EC}{\ensuremath {\mathrm {EC}}}
\newcommand {\CAR}{\ensuremath {\mathrm {CAR}}}
\newcommand {\ARAR}{\ensuremath {\mathrm {AR}\text{-}{\mathrm
      {AR}}}}
\newcommand {\AR}{\ensuremath {\mathrm {AR}}}
\newcommand {\ARbar}{\ensuremath {\overline{\mathrm {AR}}}}

\begin{abstract} 
Predictions are established for linear differential current-current
cross-correlations $dS_{a,b}/dV$ in a symmetrically biased three-terminal
normal metal-superconductor-normal metal (NSN) device. Highly
transparent contacts turn out to be especially interesting because they
feature positive $dS_{a,b}/dV$. At high transparency, processes based on Crossed
Andreev Reflection (CAR) contribute only negligibly to the current and to
$dS_{a,b}/dV$. Under these circumstances, current-current cross-correlations
can be plausibly interpreted as a coherent coupling between the two NS
interfaces in the form of {\it synchronized Andreev and inverse Andreev}
reflections ({\ARARbar}), corresponding to the process where a pair of
electron-like quasi-particles and a pair of hole-like quasi-particles arrive
from the normal electrodes and annihilate in the superconductor.
Hence,
positive $dS_{a,b}/dV$ does not automatically imply {\CAR}. For tunnel
contacts, $dS_{a,b}/dV$ is positive because of {\CAR}. In between these two
extremities, at intermediate transparencies, $dS_{a,b}/dV$ is negative because
both processes which cause positive correlations, occur only
with small amplitude. 
We use scattering theory to obtain analytic expressions for current and noise,
and microscopic calculation using a tight binding model in order to obtain a
clear interpretation of the physical processes.
\end{abstract}
\maketitle

\section{Introduction}
Beautiful experiments on transport and noise in normal metal-superconducting (NS)
hybrids allow probing the microscopic physics associated to the
superconducting condensate and quasi-particles. For instance, the
superconducting gap $\Delta$ was revealed by tunnel spectroscopy on a normal
metal-insulator-superconductor tunnel junction. In NS structures with highly
transparent contacts, Andreev reflection \cite{Andreev} is the phenomenon by
which pairs of electron-like quasi-particles from the normal electrode N can enter the
superconductor S and join the condensate. Additional processes appear in
N$_a$SN$_b$ structures with two normal electrodes N$_a$ and N$_b$: an electron
coming from N$_a$ may be transmitted as an electron into N$_b$
(elastic cotunneling, {\EC}), or it may be transmitted as a hole into 
N$_b$ (crossed Andreev reflection,
{\CAR}\cite{Byers,Deutscher,Choi,Falci,Melin-Feinberg-PRB,Melin-Duhot-EPJB,Melin-wl,Levy-Yeyati,Golubev,Kalenkov,Lambert1,Buttiker,Sols,Yamashita,Morten,Golubov,Zaikin,Koltai,Bednorz,Giazotto,Melin-Madrid,Duhot,Samuelsson,ref5,ref6-7,ref9,Bignon,Melin-Martin,Faoro,Russo,Cadden,Schonenberger,Chandra2,Beckmann,Schonenberger2}). The
amplitudes of these two processes decrease exponentially with a characteristic
length scale: the coherence length $\xi$, which is inverse proportional to the
energy gap $\Delta$ in the ballistic limit. Therefore, three-terminal
nanoscale devices with distance $R \agt \xi$ between the contacts are
especially interesting: conductance and noise experiments on them probe both
the condensate and the quasi-particle states.

Concerning current-current correlations, theorists have envisioned two kinds
of experiments for the long term: using entanglement in quantum information
devices, and testing entanglement in the electronic Einstein-Podolsky-Rosen
(EPR) experiment.\cite{Choi,Martin-Lesovik,Faoro} The EPR experiment is not
considered here, but instead the basic problem of current-current
cross-correlations \cite{Buttiker-revue} in NSN structures is addressed. Some
experiments based on NISIN structures have been
reported recently.\cite{Chandra-noise} Our task here is not to understand the
tunnel limit, where an insulating oxide layer I is inserted in between the
normal and the superconducting electrodes, but the opposite limit of highly
transparent interfaces where Coulomb interactions \cite{Levy-Yeyati} are not
expected to play a predominant role.

Based on the limiting case of tunnel contacts,\cite{Falci,Bignon} one may
erroneously conclude that positive differential current-current
cross-correlations $dS_{a,b}/dV$ are equivalent to {\CAR}. However, this is not the case
because, as we show, positive $dS_{a,b}/dV$ can well be obtained in the
absence of {\CAR}.

Current-current cross-correlations are negative \cite{Buttiker-prb,exp1,exp2}
for non-interacting fermions. A flux of
bosons leads to positive cross-correlations \cite{HBT} and negative cross-correlations are found for bosons
impinging one by one onto a beam splitter.\cite{Grangier} Cross-correlations
can be positive in interacting fermionic systems,
\cite{ref12,ref13,ref14,ref15,ref16,ref17,ref18} as well as in multiterminal
NS
structures.\cite{Duhot,Samuelsson,ref5,ref6-7,ref9,Bignon,Melin-Martin,Morten,Bednorz,Faoro}

The recent experiment \cite{Chandra-noise} and other experiments under way
involve normal electrodes separately connected to a superconductor,
\cite{Melin-Martin} with a geometry similar to that considered in the following.
Not only the noise can be evaluated in various set-ups, but also the full
histogram of the charge transmitted in a given time interval.
\cite{Bednorz,Morten,Faoro} In addition to these set-ups, relevant information
is also
obtained from ``zero dimensional'' chaotic cavities in contact with a
superconductor, in connection with the possibility to observe positive current-current
cross-correlations due to {\CAR}. \cite{Samuelsson} 

In what follows, attractive interaction binding pairs of electron-like
quasi-particles is present
everywhere in the superconducting region. Two electron-like quasi-particles of a pair injected
into the superconducting region remain glued by the BCS mean field
interaction, in contrast with the dissociation of a Cooper pair entering a
chaotic cavity (see Ref.~\onlinecite{Samuelsson}). The physics behind
current-current cross-correlations in the NSN structure with highly
transparent contacts considered here was not really elucidated in
Ref.~\onlinecite{Melin-Martin}, in spite of the important observation that
$dS_{a,b}/dV$ is positive but the non-local conductance is negative at low
bias. Unusual properties can be realized with the
following experimental conditions: First it is assumed that the same voltage 
$V_a=V_b\equiv V$ is applied on the two normal electrodes N$_a$ and N$_b$.
Second the temperature is very low,
and third high values of interface transparencies are used. Direct electron
transmission is Pauli blocked at zero
temperature because of the same voltage applied on the two
normal electrodes. On the other hand, crossed Andreev
reflection is strongly reduced at high transmission, regardless of the applied
voltages $V_a$ and $V_b$. These two features of {\EC} and {\CAR} are shown to be a
sufficient condition for deducing $dS_{a,b}/dV>0$, not due to {\CAR}, but due to
what is called here {\it synchronized Andreev reflection and inverse Andreev
  reflection} ({\ARARbar}). The positive current-current correlations at high
transparency are not in disagreement with what is found in
Ref.~\onlinecite{Samuelsson} for strong coupling to the superconductor, where
a gap is induced in a chaotic cavity. The cross-over
between high values of interface transparency and tunnel contacts will also be
investigated: {\CAR} has a dominant contribution to $dS_{a,b}/dV$ for tunnel
contacts, and $dS_{a,b}/dV$ is negative at the cross-over for intermediate
values of interface transparency, where {\ARARbar} is suppressed.

The article is divided into two independent main sections where
current-current correlations are evaluated on the basis of (i) the scattering
approach for a homogeneous superconducting gap (see Sec.~\ref{sec:scattering})
and of (ii) microscopic calculations taking into account the strong inverse
proximity effect (see Sec.~\ref{sec:micro}). Technically, these two sections
rely on different approaches and are based on different assumptions. Overall
agreement between the two calculations is obtained. Final remarks are provided
in Sec.~\ref{sec:conclu}. Technical details are as much as possible left for
Appendices. The article starts with a preliminary section containing
definitions and a summary of the main results.

\begin{figure}
\centerline{\includegraphics[width=0.9\columnwidth]{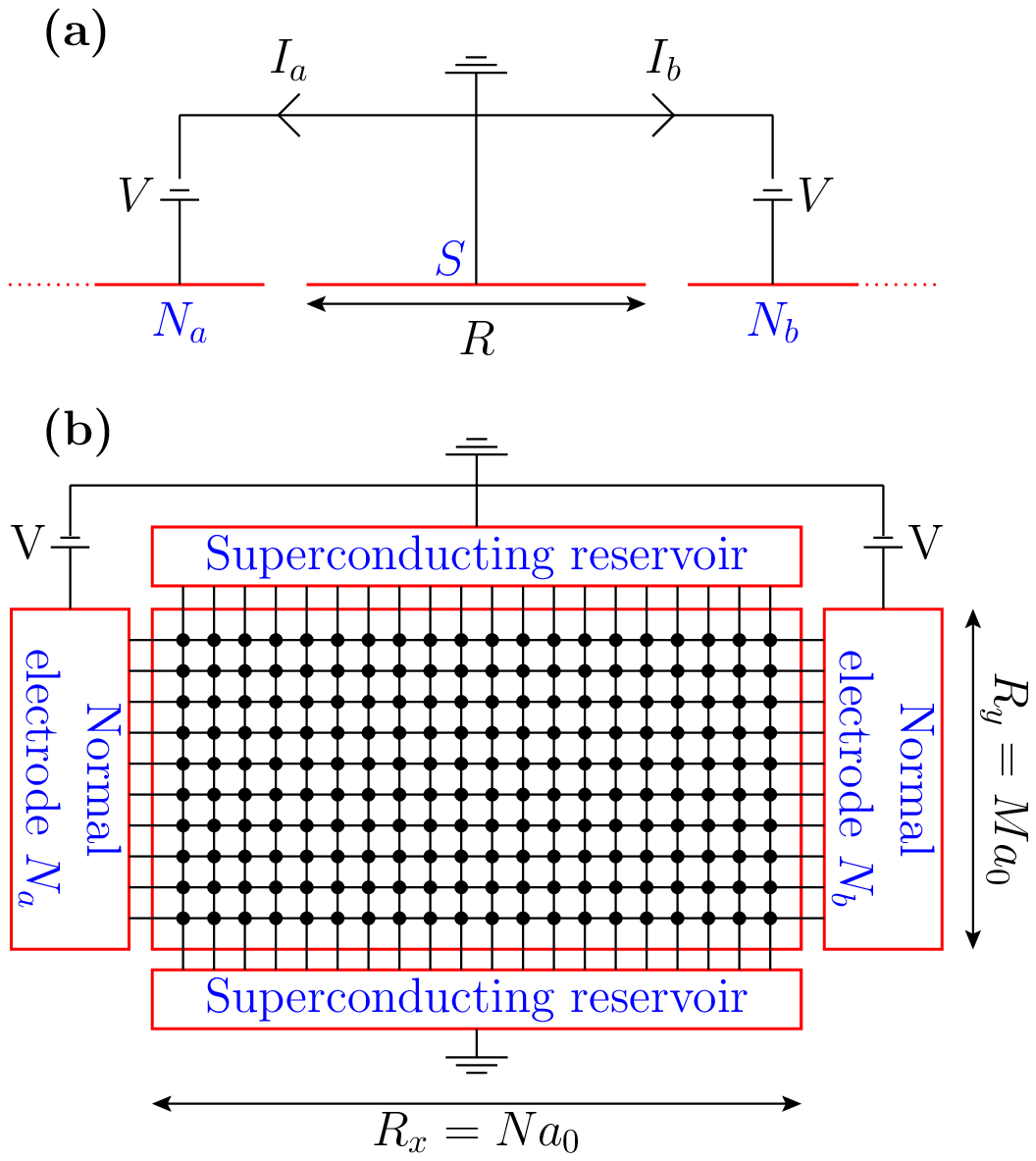}}
\caption{(Color online). 
The two models studied in this paper: Panel (a) shows
the one-dimensional geometry used in the BTK calculation, which is a simplified
description for a NSN structure in three dimensions. The central
superconducting electrode has length $R$ and the superconducting gap
$\Delta_0$ is uniform. Panel (b) shows a representation of a three-terminal device
described by a tight-binding Hamiltonian. The two superconducting reservoirs
have the same phase and thus constitute a single terminal. 
\label{fig:schema3}
}
\end{figure}

\section{Preliminaries}
\label{sec:hamil}

\subsection{Current, noise and current-current correlations}
\label{sec:intro-H}
We start with general definitions of current and current-current
cross-correlations. The geometry of the considered set-up is shown in
Fig.~\ref{fig:schema3}. The
central island is superconducting, and it is connected by highly transparent
contacts to the two superconducting reservoirs on top and bottom. The
superconductor S is made of the central island and of the two reservoirs.

The operator giving the current flowing at time $t$ from
the normal electrode N$_a$ to the superconducting island S at the N$_a$S
interface [see Fig.~\ref{fig:schema3}(b)] takes the form
\begin{align}
\label{eq:Ia}
\hat{I}_{a,\alpha}(t)&= \sum_{\sigma,n=1}^M \left(
t_{a_n,\alpha_n} c_{\alpha_n,\sigma}^+(t)
c_{a_n,\sigma}(t) \right.\\&\left.+
t_{\alpha_n,a_n} c_{a_n,\sigma}^+(t)c_{\alpha_n,\sigma}(t)\right)
\nonumber
,
\end{align}
where $\sigma$ is the projection of the spin on the quantization axis and the
sum over $n$ runs over the $M$ tight-binding sites describing the
interface. Tight-binding sites on the normal side of the interface N$_a$S are
labeled by $a_n$, and their counterparts in the superconducting electrode are
labeled by $\alpha_n$. The hopping amplitudes between electrode N$_a$ and the
superconductor are denoted by $t_{a_n,\alpha_n}$ and $t_{\alpha_n,a_n}$. One
has $t_{a_n,\alpha_n}=t_{\alpha_n,a_n}\equiv t_a$ in the absence of a magnetic
field. The average current $I_a\equiv\langle\hat{I}_{a,\alpha}(t)\rangle $
is the expectation value of the current operator given by Eq.~\eqref{eq:Ia}.

Current-current auto-correlations in electrode N$_a$ are given by
\begin{equation}
\label{eq:Saa1}
S_{a,a}(t')=\langle \delta \hat{I}_a(t+t') \delta \hat{I}_a(t) \rangle +
\langle \delta \hat{I}_a(t) \delta \hat{I}_a(t+t') \rangle
, 
\end{equation}
with $\delta \hat{I}_a(t)=\hat{I}_a(t)-\langle
\hat{I}_a(t) \rangle$. In the absence of ac-excitations, the average current
given by Eq.~\eqref{eq:Ia} is time-independent and the auto-correlations
$S_{a,a}(t')$ given by Eq.~\eqref{eq:Saa1} depend only on the difference $t'$
of the time arguments.

Similarly, current-current cross-correlations between the electrodes N$_a$ and
N$_b$ are given by
\begin{equation}
\label{eq:Sab0}
S_{a,b}(t')=\langle \delta \hat{I}_a(t+t')\delta \hat{I}_b(t) \rangle
+ \langle \delta \hat{I}_b(t)\delta \hat{I}_a(t+t') \rangle
,
\end{equation}
where $\hat{I}_b$ describes the current at the interface with the normal
electrode N$_b$. Zero-frequency auto-correlations ($S_{a,a}$) and
cross-correlations ($S_{a,b}$) are defined as the integral over $t'$ of
$S_{a,a}(t')$ and of $S_{a,b}(t')$. The definition used here for the Fano
factor is as follows: $F_{a,a}=S_{a,a}/2e I_a$ and $F_{a,b}=S_{a,b}/2e
\sqrt{I_aI_b}$, with $I_a=I_b$. With this definition, the Schottky formula
leads to $F= q^*/e$ for a Poisson process transmitting a charge~$q^*$.

\begin{figure}
\centerline{\includegraphics[width=0.9\columnwidth]{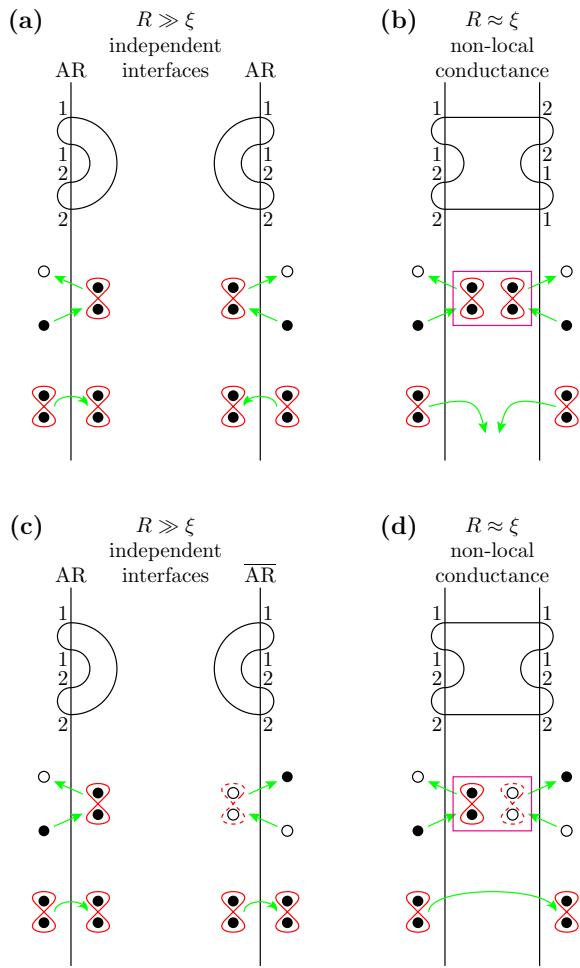}}
\caption{(Color online). 
Two independent local {\AR} processes are shown on panel (a), for two interfaces
separated by a distance $R$ much larger than the coherence length $\xi$. The
labels 1 and 2 on the diagram correspond to electrons and holes
respectively. For each {\AR} process, an electron impinging from the normal
electrode onto the interface is converted into a hole, and a pair is transmitted
into the superconductor. If $R\sim \xi$, the two {\AR} processes can be coupled
coherently by non-local propagation in the superconductor [see panel (b)], with
a quartet as an intermediate state and penetration of a charge $4e$ into the
superconductor, which is qualitatively equivalent to double CAR. Panel
(c) shows independent {\AR} and {\ARbar} processes at the two interfaces,
supposed to be far apart. Panel (d) shows {\ARARbar} for $R\sim\xi$, which is
qualitatively equivalent to double {\EC}.
\label{fig:ARAR-schema}
}
\end{figure}
\subsection{Known results and open questions}
\label{sec:open-questions}

A few facts related to the non-local conductance for arbitrary values of
interface transparency are already known. Only local Andreev reflection {\AR}
and local inverse Andreev reflection {\ARbar} come into account for contacts
separated by a distance much larger than the coherence length [see
Figs.~\ref{fig:ARAR-schema}(a) and (c)]. Local {\AR} means that an electron is
converted into a hole and a pair of electron-like quasi-particles is transmitted
into the superconductor, and local {\ARbar} means that a hole is converted into
an electron and a pair of hole-like quasi-particles is transmitted into the
superconductor, which annihilates a Cooper pair. Local {\AR} and {\ARbar}
contribute to transport if the separation between the interfaces is
much larger than $\xi$, and in two-terminal configurations where the
superconductor is not connected to ground.

Other quantum processes appear in a three-terminal configuration if
the distance between the contacts is comparable to the coherence length:
{\CAR} and {\EC}. However, non-standard types of ``non-local'' processes can
be also obtained by merging {\AR} at the interface N$_a$S to {\AR} at the
interface SN$_b$, forming what is called here {\ARAR} [see
Fig.~\ref{fig:ARAR-schema}(b)]. Physically, {\ARAR} would correspond to the
synchronized transmission of two pairs from electrodes N$_a$ and N$_b$ into
the superconductor, which can be seen as double CAR. However, this
process does not contribute to non-local transport at zero temperature.
Conversely, {\AR} at interface N$_a$S might be associated to {\ARbar} at
interface SN$_b$, leading to {\ARARbar}. The corresponding non-local
resistance is independent on the value of interface transparency in the
tunnel limit.
\cite{Melin-Duhot-EPJB,Melin-wl} Qualitatively, {\ARARbar} can also be seen
as double {\EC}.

The non-standard non-local process {\ARARbar} involving pairs appears naturally
when expanding diagrammatically \cite{Melin-wl} the non-local conductance to
order $t^8$, with $t$ the hopping amplitude at the interfaces. As it is shown
below, {\ARARbar} plays a central role in understanding the positive
\cite{Melin-Martin} current-current cross-correlations for highly transparent
contacts, in a regime which is not described by perturbation theory in $t$.

A very recent preprint \cite{Che} points out the possibility of ``synchronized
Andreev transmission'' in the current-voltage characteristics of a SNS
junction array. Synchronization manifests itself in this work as specific
features in the current-voltage characteristics of the two-terminal SNS
junction array. We arrive here at the conclusion that synchronization of Andreev
processes is also possible if the separation between the NS interfaces is
comparable to the coherence length. The dominant channel {\ARARbar} is shown
to result in positive current-current cross-correlations.

As mentioned in the introduction, experiments on current-current
cross-correlations in NSN structures have already started.\cite{Chandra-noise}
A few basic questions regarding current-current cross-correlations for highly
transparent contacts have not yet received a satisfactory explanation. 

First what is the physics behind the positive\cite{Melin-Martin} linear
differential current-current correlations $dS_{a,b}/dV$ for a highly
transparent NSN beam splitter? It is shown that $dS_{a,b}/dV>0$ is not an
evidence for {\CAR} (which would prevail \cite{Bignon} for tunnel contacts in the
absence of Coulomb interactions). An interpretation in terms of {\ARARbar} is
proposed.

Second, how do current-current cross-correlations depend on the sample
geometry? Current-current cross-correlations decay with the geometry-dependent
coherence length, as it will be obtained from microscopic calculations in
Sec.~\ref{sec:micro}.

Third, what is the value of cross-correlations at intermediate transparencies?
Experimentalists can realize tunnel or highly transparent contacts, by
oxidizing or not the sample during fabrication. Intermediate values of
interface transparency are more difficult to control but it is nevertheless
useful to quantify how ``perfectly transparent'' the NS contacts should be in
order to obtain {\ARARbar}. A cross-over from positive to negative $dS_{a,b}/dV$ is
found in BTK calculations as the normal state transmission coefficient $T_N$ is
reduced below a value typically of order $T_N\sim 1/2$. 

Fourth, do the predictions established with a ballistic superconductor hold
also for a disordered superconductor? It is shown at the end of
Sec.~\ref{sec:micro} that, for strong inverse proximity effect, {\ARARbar} is
responsible for positive cross-correlations also in the case of a disordered
superconductor in the regime where the elastic mean free path is shorter than
the coherence length.

\section{Homogeneous superconducting gap: BTK calculation}
\label{sec:scattering}
The BTK approach \cite{BTK} allows calculations of the current-voltage
characteristics of a NS point contact with arbitrary interfacial scattering
potential. It was first generalized in Ref.~\onlinecite{Melin-Duhot-EPJB} and
later in Ref.~\onlinecite{Zaikin} to the case of non-local transport. Useful
physical informations can be obtained even though the gap is not
self-consistent in the BTK calculation. We will consider a one-dimensional
geometry, as shown in Fig.~\ref{fig:schema3}(a).

The current $I_i$ at the interface with the normal electrode N$_i$, and the
current-current correlations $S_{i,j}$ between the two normal electrodes N$_i$
and N$_j$ are expressed\cite{ref5} in terms of the scattering matrix $s$ as
\begin{align}
\label{eq:couranti}
I_i&=\frac{e}{h}\int d\omega
\sum_{j,\alpha,\beta}\mbox{sgn}(\alpha)
\left[\delta_{i,j}\delta_{\alpha,\beta}-
\left|s_{i,j}^{\alpha,\beta}\right|^2\right]
f_{\beta}(\omega)\\
\label{eq:Sij}
S_{i,j}&=\frac{2e^2}{h}\int d\omega
\sum_{k,l,\alpha,\beta,\gamma,\delta} \mbox{sgn}(\alpha)
\mbox{sgn}(\beta)\\
\nonumber
&A_{k\gamma,l\delta}(i,\alpha,\omega)
A_{l\delta,k\gamma}(j,\beta,\omega) f_{\gamma}(\omega)
\left[1-f_{\delta}(\omega)\right]
,
\end{align}
with
\begin{equation}
A_{k\gamma,l\delta}(i,\alpha,\omega)=\delta_{i,k}
\delta_{i,l}\delta_{\alpha,\gamma}\delta_{\alpha,\delta}
-s_{i,k}^{\alpha,\gamma\dagger} s_{i,l}^{\alpha,\delta}
.
\end{equation}
Latin labels $i,j,k,l$ run over $a$,$b$, referring to the two normal electrodes
N$_a$ and N$_b$. Greek labels $\alpha,\beta,\gamma,\delta$ denote electrons or
holes in the superconductor. The notation $f_{e}(\omega)=\theta(eV-\omega)$
stands for the distribution function of electrons at zero temperature, and
$f_{h}(\omega)=\theta(-eV-\omega)$ is the one of holes, where $\theta(x)$ is the
Heaviside step-function. In Eqs.~\eqref{eq:couranti}
and~\eqref{eq:Sij},  $\mbox{sgn}(\alpha)=+1$ if $\alpha=e$, and
$\mbox{sgn}(\alpha)=-1$ if $\alpha=h$.

\begin{figure*}
\centerline{\includegraphics[width=0.9\textwidth]{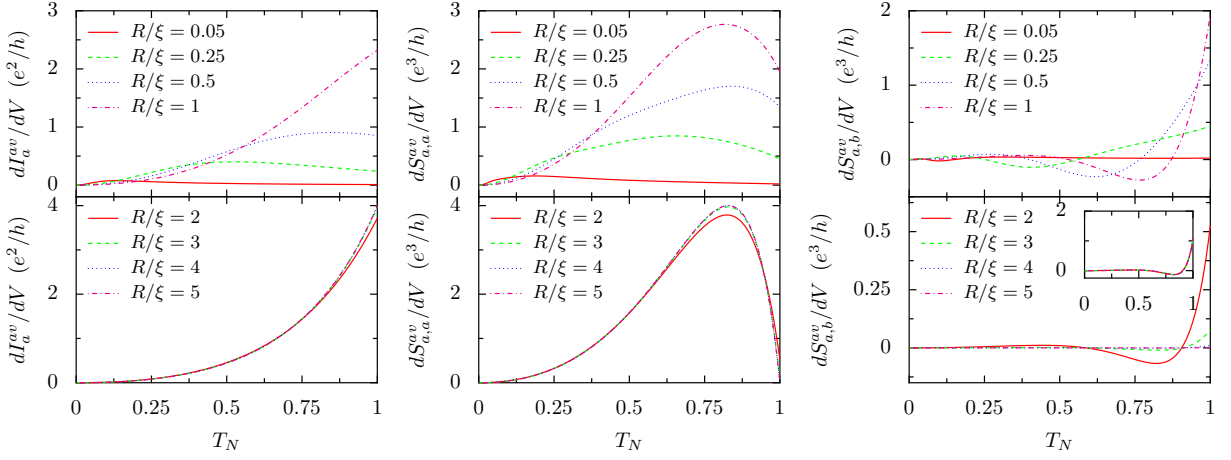}}
\caption{(Color online). 
Variations of the linear differential conductance [left panels, in
units of $e^2/h$], linear differential auto-correlations [center panels,
in units of $e^3/h$] and linear differential cross-correlations [right panels,
in units of $e^3/h$] as a function of
the normal interface transparency $T_N$, for the values of $R/\xi$ indicated
in the figures. A single-channel one-dimensional BTK calculation is used [see
Fig.~\ref{fig:schema3}(a)]. The lower Panels show that $dI_a^{av}/dV$ and
$dS_{a,a}^{av}/dV$ are almost independent on $R/\xi$ for $R/\xi\agt 1$, and that
$dS_{a,b}^{av}/dV$ becomes very small as $R/\xi$ increases above $\sim 1$. 
The insert of the bottom right panel 
shows the variations of the normalized current-current cross-correlations $[d
  S_{a,b}^{av}/dV (T_N)]/[d S_{a,b^{av}}/dV   (T_N=1)]$. The data corresponding to
$R/\xi=2,\,3,\,4,\,5$ superimpose after this rescaling. 
\label{fig:Saa-Sab-vs-TN}
}
\end{figure*}

\begin{figure*}
\centerline{\includegraphics[width=0.7\textwidth]{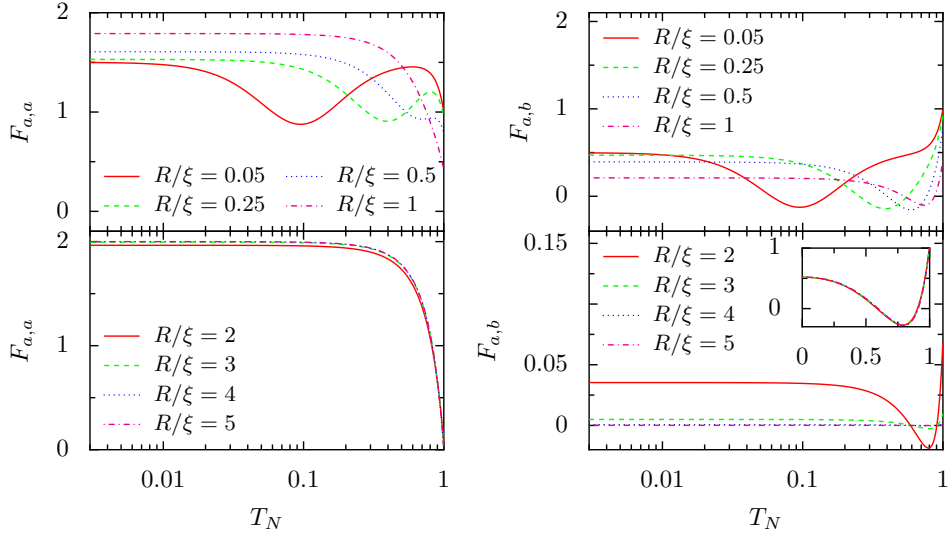}}
\caption{(Color online). 
Variations of the Fano factors $F_{a,a}=S_{a,a}^{av}/2eI_a^{av}$ [top panels]
and $F_{a,b}=S_{a,b}^{av}/2eI_a^{av}$ [bottom panels] as a function
of the normal interface transparency $T_N$, for the values of $R/\xi$
indicated in the figures. A single-channel one-dimensional BTK calculation is
used [see Fig.~\ref{fig:schema3}(a)]. The ratio between the gap $\Delta_0$ and the
Fermi energy $\epsilon_F$ is $\Delta_0/\epsilon_F=10^{-4}$ in this simulation.
The insert of bottom right panel shows the variations of the normalized crossed Fano
factor $[F_{a,b} (T_N)]/[F_{a,b} (T_N=1)]$. The data corresponding to
$R/\xi=2,\,3,\,4,\,5$ superimpose after rescaling.
\label{fig:Faa-Fab-vs-TN}
}
\end{figure*}

The elements of the $s$-matrix are evaluated from the BTK approach (see
Appendix~\ref{app:details-scat}) for a one-dimensional N$_a$SN$_b$ junction
[see Fig.~\ref{fig:schema3}(a)]. Step-function variation of the superconducting
gap at the interfaces is assumed. A repulsive scattering potential
$V(x)=H\left[\delta(x)+\delta(x-R)\right]$ is introduced at the
interfaces. The transparency of the interfaces is related to the BTK parameter
$Z=H/\hbar v_F$, with the Fermi velocity $v_F$. The interface transparency is
characterized by the value of the normal state transmission coefficient
$T_N=1/(1+Z^2)$. Highly transparent contacts correspond to $T_N=1$, and tunnel
contacts correspond to $T_N\ll 1$.

In the one-dimensional model considered in this section, current and noise
are highly sensitive to the length $R$ of the superconducting region: they
oscillate as a function of $R$ with a period equal to the Fermi wave-length
$\lambda_F\ll R$. These oscillations can be interpreted as Friedel oscillations
where the contacts with the normal electrodes play the role of impurities. In a
more realistic higher-dimensional model with more than one transmission mode,
the oscillations in the different modes are independent and thus are averaged
out. In order to simulate qualitatively multi-dimensional behavior with a
one-dimensional system, current and noise are averaged over one oscillation
period: \cite{Falci,Melin-Feinberg-PRB,Melin-Martin}
\begin{align}
\label{eq:Iav}
I_i^{av}(R)&= \frac{1}{\lambda_F}
\int_{R-\lambda_F/2}^{R+\lambda_F/2} dr \kern2pt I_i(r)\,,\\
S_{i,j}^{av}(R)&=  \frac{1}{\lambda_F}
\int_{R-\lambda_F/2}^{R+\lambda_F/2} dr \kern2pt S_{i,j}(r)
\label{eq:Sav}
.
\end{align}
In the studied limit of small applied voltage  $eV\ll\Delta$, the
energy-dependence of the scattering matrix elements can be neglected and
$s_{i,j}^{\alpha,\beta}(\omega)\approx s_{i,j}^{\alpha,\beta}(\omega=0)$. 
Using this approximation, it is possible to perform these integrals
analytically.

In this limit, the current $I_i$ through the normal lead N$_i$ is obtained as
\begin{equation}
\label{eq:I_analytic}
I_i^{av}= 
\frac{ 2 \sqrt{2} e^2V T_N^2(2-T_N)\COSH\SINH^2 }
     {h\left[2(2-T_N)^2\COSH^2-8(1-T_N)\right]^{\frac 3 2}}\,.
\end{equation}
The local current-current correlations $S_{ii}$ at the interface N$_i$S and the
current-current cross-correlations $S_{ij}$ between the interface N$_i$S and
the interface SN$_j$ are 
\begin{align}
\label{eq:Saa_analytic}
&\begin{aligned}
S_{ii}^{av} =&
\frac { 64 \sqrt{2} e^3V T_N^2 \SINH^2 \SECH }
      {h (2-T_N)^3 \left[2(2-T_N)^2\COSH^2-8(1-T_N)\right]^{\frac 7 2} }\\
  &\Big[-128(1-T_N)^4 +2(2-T_N)^2(1-T_N)(96-T_N\\
  &\times(192-T_N(116-T_N(20-3T_N))))\COSH^2\\
  &-(2-T_N)^4(72-T_N(144-T_N(82-T_N\\
  &\times(10+T_N))))\COSH^4\\
  &+8(2-T_N)^6(1-T_N)\COSH^6\Big]
\end{aligned}\\
\intertext{and}
\label{eq:Sab_analytic}
&\begin{aligned}
S_{i\neq j}^{av} =& 
\frac { 64 \sqrt{2} e^3V T_N^2 \SINH^2 \SECH }
      {h (2-T_N)^3 \left[2(2-T_N)^2\COSH^2-8(1-T_N)\right]^{\frac 7 2} }\\
   &\Big[128(1-T_N)^4 -2(2-T_N)^2(1-T_N)(2+T_N)\\
   &\times(2-3T_N)(8-(8-T_N)T_N)\COSH^2\\
   &+(2-T_N)^4(8-T_N(16+T_N(2-T_N(10+T_N))))\\
   &\times\COSH^4\Big]\,.
\end{aligned}
\end{align}

The linear conductance $dI_a^{av}/dV$, the auto-correlations
$dS_{a,a}^{av}/dV$ and the cross-correlations $dS_{a,b}^{av}/dV$ given by these
formulas are
shown in Fig.~\ref{fig:Saa-Sab-vs-TN} as a function of the normal transmission
coefficient $T_N$, for different values of the distance between the
contacts. As it is expected, the conductance increases with interface
transparency for $R/\xi\agt 1$. The conductance depends on the ratio $R/\xi$
while $R$ is smaller than the coherence length $\xi$, but it does almost not
vary as $R$ is increased above $\xi$. For $R/\xi<1$, the conductance  is
non-monotonous when plotted as a function of $T_N$ (see
Fig.~\ref{fig:Saa-Sab-vs-TN} top left). An explanation for the non-monotonous
behavior is the enhanced transmission due to the finite size of the
superconductor.\cite{Beenakker}

Starting from tunnel contacts, the linear differential auto-correlation
$dS_{a,a}^{av}/dV$  first increases with interface transparency as larger
current leads to larger noise. The differential noise reaches a maximum and
almost vanishes for perfect transparency if $R\agt \xi$, as it is expected for a
single NS junction.\cite{Buttiker-revue} Differential current-current
cross-correlations $dS_{a,b}^{av}/dV$ are positive for $R/\xi\alt 1$ in the
extreme cases of very high and very low interface transparency, and take
negative values in between, as shown in Figs.~\ref{fig:Saa-Sab-vs-TN}(right),
and in the insert of Fig.~\ref{fig:Saa-Sab-vs-TN}(bottom right).

The Fano factors $F_{a,a}=S_{a,a}^{av}/2eI_a^{av}$ and
$F_{a,b}=S_{a,b}^{av}/2eI_a^{av}$ correspond to the noise normalized to the
current, which allows to get rid of the trivial effect that, at low
transparency, the noise increases when the current increases. The variations of
$F_{a,a}$ and $F_{a,b}$ feature local minima at intermediate values of
$T_N$. In the insert of Fig.~\ref{fig:Faa-Fab-vs-TN}(bottom right), the Fano factor
$F_{a,b}$ is shown for different $R/\xi\agt 1$. As it can be seen, for 
$R/\xi\agt 1$ the Fano factor is independent of $R/\xi$ after normalizing to its
value for $T_N=1$. The Fano factor $F_{a,b}$ is positive outside the region of
the minima. For $R/\xi\agt 1$, $F_{a,a}$ takes the value $F_{a,a}\simeq 2$ for
$T_N\ll 1$, and the value $F_{a,a}\simeq 0$ for $T_N=1$.

We first make some remarks in order to confirm the validity of our calculation.
Expected behavior is recovered in some known limiting cases: 

(i) For tunnel contacts ($T_N\ll 1$), the Fano factors are given by
$F_{a,a}\approx 2-\frac 1 2 \SECH$ and $F_{a,b}\approx \frac 1 2 \SECH$, which is in agreement
with the expected limiting values $F_{a,a}=2$ and $F_{a,b}=0$ for
$R/\xi\rightarrow\infty$. This corresponds to the doubling of the effective
charge for Andreev reflection at a single NS interface in the tunnel
limit.\cite{Khlus,Jehl} 

(ii) For highly transparent interfaces ($T_N\approx1$), the Fano factors vanish
exponentially with increasing $R/\xi$ as $F_{a,a}=F_{a,b}\approx \SECH$, as it
is expected for a single highly transparent NS interface. 

(iii) $dS_{a,b}^{av}/dV$ and $F_{a,b}$ are positive and very small for
$R/\xi\agt 1$ and $T_N\simeq 1$, in agreement with Ref.~\onlinecite{Bignon}.
Only {\CAR} contributes to current-current cross-correlations for $T_N\ll 1$.

(iv) As it can be seen in
Fig.~\ref{fig:Saa-Sab-vs-TN}(top left), the linear conductance $dI_a^{av}/dV$ is
suppressed for $R/\xi\alt 1$ because the number of normal states within the
gap energy is $\sim R/\xi$ for this geometry. For a three-dimensional grain it
is proportional to $\sim k_F^2 R^3/\xi$, which can be much larger than unity
even for $R/\xi\alt 1$. The suppression of Andreev processes at high
transparency for $R/\xi\ll 1$ is thus not expected to occur in the case of a
three-dimensional superconducting grain. 

The Schottky limit is realized for low values of interface transparencies, which
lead to $\langle (\delta \hat{I}_a-\delta \hat{I}_b)^2\rangle_{av}=4eI_a^{av}$.
On the other hand $\langle (\delta \hat{I}_a+\delta
\hat{I}_b)^2\rangle_{av}=4e(I_a^{av}+I_b^{av})$. One concludes that
$F_{a,a}\simeq 3/2$ and $F_{a,b}\simeq 1/2$, in agreement with
the plateau obtained in Figs.~\ref{fig:Faa-Fab-vs-TN}(top) for
the dependence on $T_N$ of the Fano factor. Eqs.~\eqref{eq:I_analytic},
\eqref{eq:Saa_analytic} and \eqref{eq:Sab_analytic} reproduce these values in
the limit $R/\xi\rightarrow 0$ and $T_N\ll 1$.

For highly transmitting interfaces $T_N=1$, we obtain $F_{a,a}=F_{a,b}$. These
values can be confirmed by a more simple calculation in the limit $R/\xi\ll 1$
(see Appendix~\ref{app:FaaFab2e}).  The identity $F_{a,a}=F_{a,b}$ implies that
$\hat{I}_a^{av}-\hat{I}_b^{av}$ is noiseless for $T_N=1$, independent of
$R/\xi$:
\begin{equation}
\label{eq:totoa}
\begin{aligned}
&\int d\tau
\langle\left( \delta \hat{I}_a(t)-\delta \hat{I}_b(t)\right)
\left( \delta \hat{I}_a(t+\tau)-\delta \hat{I}_b(t+\tau)\right) \rangle_{av}\\
&=\frac{1}{2}\left(
S_{a,a}^{av}+ S_{b,b}^{av}-2S_{a,b}^{av}\right)\\
&= \frac{I_a^{av}}{2}\left(F_{a,a}+F_{b,b}-2F_{a,b}\right)=0
.
\end{aligned}
\end{equation}
By comparison, in a fermionic beam splitter with highly transparent contacts, a
charge $e$ transmitted from the source
to N$_a$ means no charge transmitted from the source to N$_b$. Thus, for
fermions, it is the sum $\hat{I}_a+\hat{I_b}$ that is noiseless. 

We can see from Fig.~\ref{fig:Saa-Sab-vs-TN} that the linear differential cross-correlations
of current noise $dS_{a,b}^{av}/dV $ are positive for $T_N\ll1$ and $T_N\simeq 1$, while
they take negative values in between these two limiting cases. For
$T_N\ll1$, the positive $dS_{a,b}^{av}/dV$ can be explained by the presence of
CAR processes. This is in agreement with perturbative calculations carried out
in Ref.~\onlinecite{Bignon}. However, for perfectly transmitting interfaces
$T_N\simeq1$, CAR processes do not occur, as the elements of the scattering
matrix describing CAR (e.g.\ $s_{a,b}^{e,h}$) equal zero for $T_N=1$ (see
Appendix~\ref{app:FaaFab2e}).  In this Appendix we obtain in the limit $R/\xi\ll1$
\begin{align}
\label{eq:identity1}
\frac{d S_{a,b}^{av}}{dV}&=8\frac{e^3}{h} 
\begin{cases}\left | s_{a,a}^{e,h}\right|^2 &\text{ for } R/\xi\ll1\\
\left | s_{a,b}^{e,e}\right|^2 =T_{\EC} &\text{ for } R/\xi\agt1\end{cases}\\
\intertext{and}
\label{eq:identity2}
T_{\CAR}&=\left|s_{a,b}^{e,h}\right|^2=0
.
\end{align}
In section~\ref{sec:micro}, we will use a microscopic model in order to obtain an
understanding of the processes contributing to the noise, and to explain
$dS_{a,b}^{av}/dV > 0$ in the absence of CAR.

To summarize this section, we have obtained analytical expressions for the
current and noise in the one-dimensional BTK model. It was shown that
$dS_{a,b}^{av}/dV>0$, and that Eq.~\eqref{eq:identity1} holds for $R/\xi\agt 1$
and for $T_N\simeq 1$. Cooper pair splitting dominates for small normal
transmission coefficient $T_N\ll 1$, while what will be interpreted in
Sec.~\ref{sec:micro} as {\ARARbar} dominates for $T_N\simeq 1$. These two
processes are suppressed for intermediate $T_N$, resulting in negative
cross-correlations in this parameter range.

\section{Microscopic calculations}
\label{sec:micro}

The current-current cross-correlations for highly transparent interfaces can be
further investigated in the two-dimensional tight-binding set-up shown in
Fig.~\ref{fig:schema3}(b).
First, using analytic calculations, we analyze the different contributions to the
noise $S_{a,b}$ and show the absence of contributions due to CAR processes. The
positive contributions are attributed to processes which we refer to as
\ARARbar{} (this notation stands for Andreev reflection and inverse Andreev
reflection), a denomination which is motivated by the microscopic analysis.
Second, numerical calculations are performed, which take into account the
inverse proximity effect by determining the gap in a self-consistent manner. In
addition, disorder will be included in the calculation.
As in the last section, we restrict our study to the case where the same
voltages $V_a=V_b=V$ are applied
on both normal leads and $V$ is small compared to the gap~$\Delta$.

The expression of current-current cross-correlations $S_{a,b}$ can be decomposed
as a sum of six contributions according to the types of transmission modes in
the superconductor:
\begin{equation}
S_{a,b} = S_{\CAR} + S_{\EC} + S_{\ARARbar} + S_{\ARAR} + S'+S_\mathrm{MIXED},
\end{equation}
where the expression and the meaning of the different contributions are provided in
Appendix~\ref{app:micro-noise}. $S_{\CAR}$ contains the noise attributed to
crossed Andreev reflections, which contains transmission modes in the
electron-hole channels. $S_{\EC}$, the noise due to elastic cotunneling, contains
transmission modes in the electron-electron or hole-hole channels. With ${\ARAR}$,
we refer to synchronous local Andreev reflections at both interfaces, while
${\ARARbar}$ links a local Andreev process at one interface to a local inverse
Andreev process at the other one (see Fig.~\ref{fig:ARAR-schema}).

\begin{figure}
\centerline{\includegraphics[width=0.9\columnwidth]{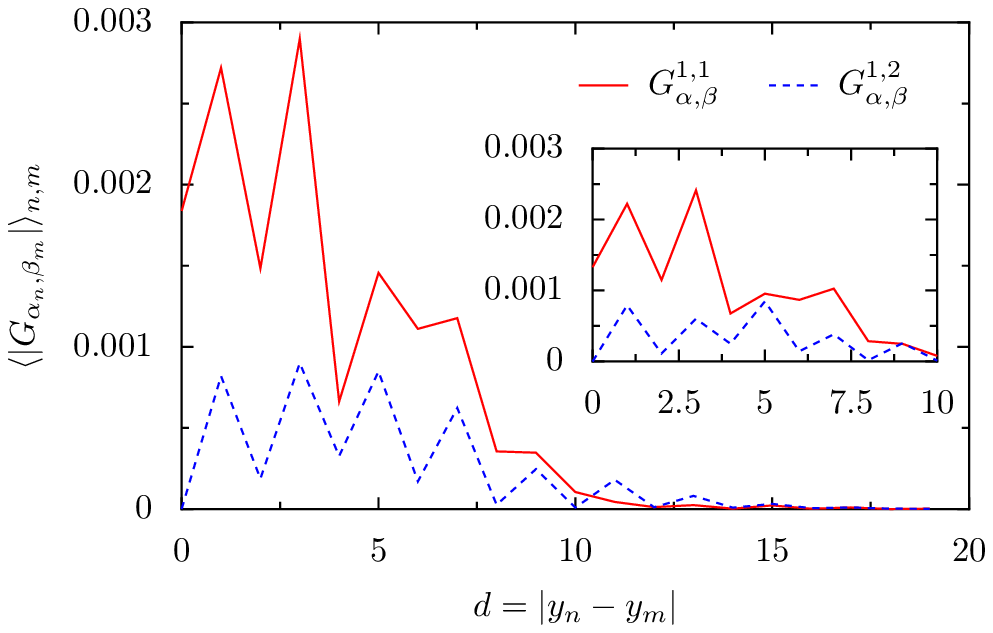}}
\caption{(Color online).
The anomalous Green function $G_{\alpha_n,\beta_m}^{1,2}$ is small
compared to the normal Green function $G_{\alpha_n,\beta_m}^{1,1}$.  As the
elements $G_{\alpha_n,\beta_m}$ (between site $n$ on the left interface and site
$m$ on the right interface) decay exponentially with $d=|y_n-y_m|$,
we have evaluated the mean values $|\langle G_{\alpha_n,\beta_m}|\rangle_{n,m}$
of the elements with the same value of $d$. The main contribution to non-local
transport comes from small values of $d$.
While the main frame shows the results obtained for a constant gap $\Delta$, the
inset shows the corresponding data with the self-consistent gap.
\label{fig:car_zero_micro}
}
\end{figure}

The Andreev-reflection is highly local\cite{Zaikin} (compare with the
BTK calculations in Sec.~\ref{sec:scattering}, where it vanishes exactly). This
motivates the assumption $|G_{\alpha,\beta}^{1,2}| \ll
|G_{\alpha,\beta}^{1,1}|$, which is confirmed by our numerical calculations also
in the presence of the inverse proximity effect (see
Fig.~\ref{fig:car_zero_micro}). 
Using this simplification, one obtains $S_{\CAR}=S_{\ARAR}=0$. The terms
$S'$ and $S_\mathrm{MIXED}$ vanish after integration over the energy $\omega$.
Thus the total current-current cross-correlation $S_{a,b}$ depends only on the
term $S_{\ARARbar}$. 

\begin{figure}
  \centerline{\includegraphics[width=0.8\columnwidth]{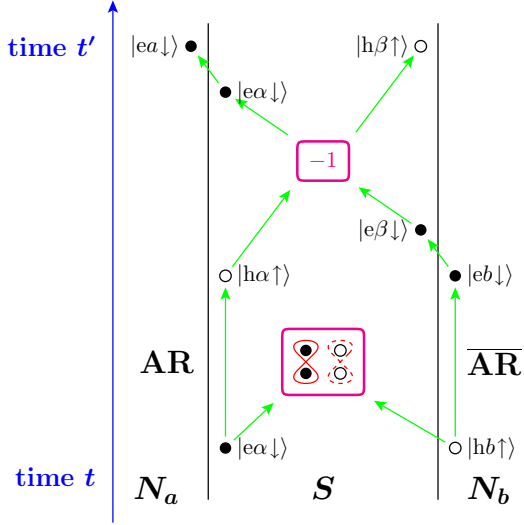}}
\caption{(Color online). 
Schematic representation of how {\ARARbar} couples to current-current
cross-correlations. The initial state consists of (i) a spin-down electron
created at $\alpha$ at time $t$, on the superconducting side of the N$_a$S
interface, and of (ii) a spin-up electron destroyed at time $t$ at b, on the
normal side of interface SN$_b$. The final state consists of a spin-down
electron destroyed at time $t'$ at a, on the normal side of interface N$_a$S,
and of a spin-up electron created at time $t'$ at $\beta$, on the
superconducting side of interface SN$_b$. To the {\ARARbar} process shown on
Fig.~\ref{fig:ARAR-schema}d, the permutation of
two fermions is added. Taking into account the resulting minus sign
leads to positive current-current cross-correlations.
\label{fig:GpmGmp}
}
\end{figure}

In order to understand what type of microscopic processes are described in
\ARARbar{}, we  start from
the formula giving the current-current cross-correlations in terms of the
Keldysh Green functions $\hat{G}^{+,-}$ and $\hat{G}^{-,+}$:
\begin{equation}
\begin{aligned}
S_{a,b}(\omega)=&\mbox{Tr}\left[ \hat{G}^{+,-}_{b,a}
  \hat{G}^{-,+}_{\alpha,\beta} + \hat{G}^{+,-}_{\beta,\alpha}
G^{-,+}_{a,b}\right.\\
&-\left.\hat{G}^{+,-}_{b,\alpha}
\hat{G}^{-,+}_{a,\beta}-
\hat{G}^{+,-}_{\beta,a}
\hat{G}^{-,+}_{\alpha,b}\right]
,
\end{aligned}
\end{equation}
where the trace is carried out over the Nambu labels and the different transmission
modes at the interfaces. It can be shown that all terms in $S_{\ARARbar}$ are
obtained from the anomalous contributions of the type
$G^{+,-,1,2} G^{-,+,2,1}$ and $G^{+,-,2,1} G^{-,+,1,2}$. Let us
consider one of these terms as an example (the same conclusions are obtained
for all terms) and suppose again that $G_{\alpha,\beta}^{A,1,2}=0$ if $\alpha$
and $\beta$ are on different interfaces (that is, crossed Andreev reflection
does not contribute), and that a symmetric bias voltage is
applied on the two normal electrodes. One has 
\begin{equation}
\begin{aligned}
&G^{+,-,2,1}_{b,a}(t,t') G^{-,+,1,2}_{\alpha,\beta}(t',t)\\
&=\langle
c_{a,\downarrow}(t')c_{b,\uparrow}(t)\rangle\langle c_{\beta,\uparrow}^+(t')
c_{\alpha,\downarrow}^+(t)\rangle
.
\end{aligned}
\end{equation}
This equation can be understood as a relation between initial and final
states, as it is shown in  Fig.~\ref{fig:GpmGmp}. In general, these states can be
connected by many different processes. However, the microscopic formula for
the current-current cross-correlations [see Eq.~\eqref{eq:SARARbar}] shows that
the initial and final states are related by an Andreev process at interface
N$_a$S, and, at the same time, by an inverse Andreev process at interface SN$_b$
[see Fig.~\ref{fig:GpmGmp}(b)]. In an Andreev process, an electron
is converted into a hole and a pair of electron-like quasi-particles is
transmitted into the superconductor. In an inverse Andreev process, a hole is
converted into an electron and a pair of hole-like quasi-particles is
transmitted into the superconductor. The pair of electron-like quasi-particles
annihilates with the pair of hole-like quasi-particles and the remaining electron and
the hole are exchanged between the two interfaces. This results in
$dS_{\ARARbar}/dV > 0$.

\begin{figure}
  \centerline{\includegraphics[width=0.9\columnwidth]{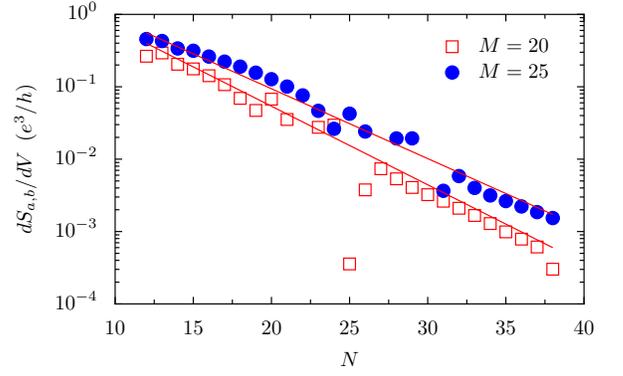}}
\caption{(Color online). 
The figure shows $dS_{a,b}/dV$ as a function of $N$ for highly transmitting
interfaces [for $N$ see Fig.~\ref{fig:schema3}(b)]. In agreement with the BTK
calculations $dS_{a,b}/dV>0$.  The exponential decay is described by the
coherence length which increases with $M$, in agreement with
Ref.~\onlinecite{Melin-Madrid}. The values $M=20$ (red squares) and $M=25$ (blue
circle) are used. Strong deviations from the exponential decay (red lines)
appear for $N\simeq M$ (see Ref.~\onlinecite{Melin-Madrid}).  The data points
are obtained from Eq.~\eqref{eq:Sab}.  The decay is $ \propto
\exp[-2Na_0/\xi(M)]$.
\label{fig:des_Sab}
}
\end{figure}

In addition to the analytic calculation, we performed numerical simulations in
order to analyze a more realistic model.
Details about the used method \cite{Melin-Madrid} are presented in
Appendix~\ref{app:micro-details}. The self-consistent simulations presented
below take into account the inverse proximity effect corresponding to the
reduction of the superconducting gap within a distance $\sim \xi$ from the
contacts. Self-consistency is equivalent to current conservation for the
electrons injected from the normal reservoirs and transmitted into the
superconducting ones.

\begin{figure}
  \centerline{\includegraphics[width=0.9\columnwidth]{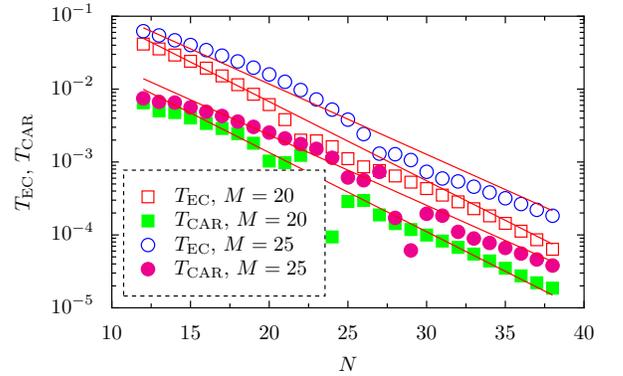}}
\caption{(Color online). 
The exponential decay of the {\EC} transmission coefficient $T_{\EC}$
[Eq.~\eqref{eq:TEC-expression}, open red squares for $M=20$ and open blue
circles for $M=25$], and of the crossed Andreev transmission coefficient
$T_{\CAR}$ [Eq.~\eqref{eq:TCAR-expression}, filled green squares for $M=20$ and
filled purple circles for $M=25$] is shown. The decay is $\propto
\exp[-2Na_0/\xi(M)]$, with $\xi(20)=8a_0$ and $\xi(25)=9a_0$.
\label{fig:des_TEC_TCAR}
}
\end{figure}

The values for the linear differential cross-correlations $dS_{a,b}/dV$, and for
the {\EC} and {\CAR} transmission coefficients $T_{\EC}$ and $T_{\CAR}$ are
plotted as functions of the length $N$ of the superconductor in
Figs.~\ref{fig:des_Sab} and \ref{fig:des_TEC_TCAR}. The notations $T_{\EC}$
and $T_{\CAR}$ refer to the transmission modes in the superconductor
(advanced-advanced or retarded-retarded modes not exchanging electrons and holes
for $T_{\EC}$, and advanced-retarded Green functions exchanging electrons and
holes for $T_{\CAR}$).
$T_{\EC}$ and $T_{\CAR}$ are given by
\begin{align}
\label{eq:TEC-expression}
T_{\EC}&= W^2 \text{Tr}\left(
G^{A,1,1}_{\alpha,\beta}G^{R,1,1}_{\beta,\alpha}
+
G^{A,2,2}_{\alpha,\beta}G^{R,2,2}_{\beta,\alpha}
\right)\\
\label{eq:TCAR-expression}
T_{\CAR}&= W^2 \text{Tr}\left(
G^{A,1,2}_{\alpha,\beta}G^{R,2,1}_{\beta,\alpha}
+
G^{A,2,1}_{\alpha,\beta}G^{R,1,2}_{\beta,\alpha}
\right)
,
\end{align}
where $W$ is the hopping amplitude in the bulk and at the interfaces, and
$\alpha_n$ and $\beta_m$ run respectively over all the sites on the
superconducting side of the N$_a$S and SN$_b$ interfaces. The notation
$G_{i,j}^{A(R),n_i,n_j}$ stands for the Nambu component $(n_i, n_j)$ of the
advanced (retarded) Green function connecting $i$ to $j$.

The dependence on $M$ of the
coherence length was already found in a previous work.\cite{Melin-Madrid} The
differential cross-correlations $dS_{a,b}/dV$ are positive, which is in
agreement with the preceding BTK calculation. Differential cross-correlations
$dS_{a,b}/dV$ show exponential decay (see Fig.~\ref{fig:des_Sab}) as a
function of $N$, because the two normal electrodes N$_a$ and N$_b$ are
coherently coupled by evanescent states in the superconductor.  
The BCS coherence length as obtained from the fits fulfills $R_x / \xi \agt 1$
in a wide range of simulation parameters. This is the range in which the
BTK calculation leads to $d S_{a,b}^{av}/dV=8\frac{e^3}{h} T_{\EC}$.

For a highly transparent NS contact, one has $G_{\alpha_n,\alpha_n}^{2,1}
G_{\beta_{m},\beta_{m}}^{1,2}\simeq 1/4W^2$ where $W$ is the hopping amplitude
in the normal and superconducting electrodes, and at the interface (see
Appendix~\ref{app:Galpha-beta}). The identity $T^{A,A}_{\ARARbar}\simeq T_{EC}$
holds if $|G_{\alpha,\beta}^{1,2}|\ll|G_{\alpha,\beta}^{1,1}|$ (see
Appendix~\ref{app:Galpha-beta}), with
\begin{equation}
T_{\ARARbar}^{A,A}=- W^2 \text{Tr}\left(
G^{A,1,1}_{\alpha,\beta}G^{A,2,2}_{\beta,\alpha}
+
G^{A,2,2}_{\alpha,\beta}G^{A,1,1}_{\beta,\alpha}
\right)
,
\end{equation}
where the superscript ``A,A'' refers to an advanced-advanced transmission
mode where electrons and hole are conserved. With these assumptions, the total
noise can be written as 
\begin{align}
\label{eq:ID1}
\frac{d S_{a,b}}{dV}&\simeq 8 \frac{e^3}{h} T^{A,A}_{\ARARbar}\,,\\
T^{A,A}_{\ARARbar}&\simeq T_{\EC}
\label{eq:ID2}
.
\end{align}

\begin{figure*}
  \centerline{\includegraphics[width=0.9\textwidth]{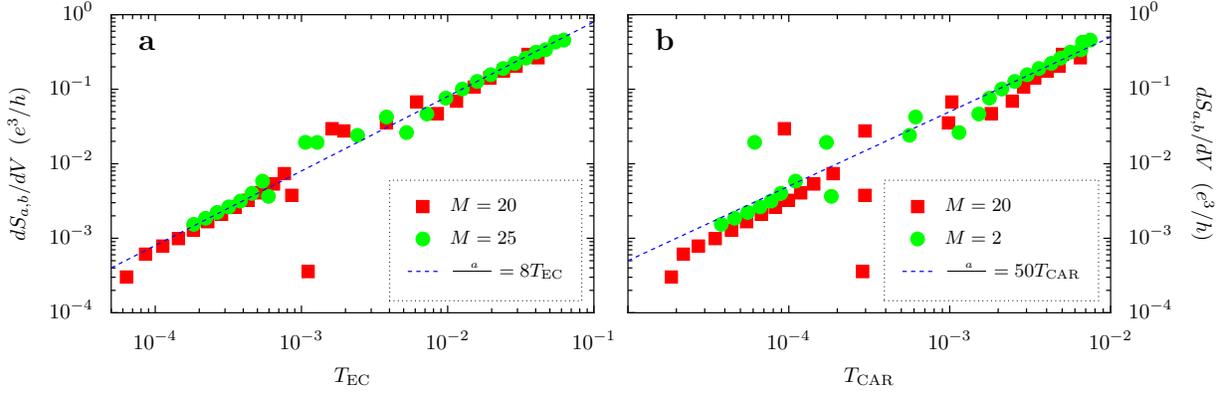}}
 \caption{(Color online). 
Relation between $dS_{a,b}/dV$ and the transmission coefficients $T_{\EC}$ and
$T_{\CAR}$: Panel (a) shows the relation between $dS_{a,b}/dV$ and $T_{\EC}$ and
panel (b) shows the relation between $dS_{a,b}/dV$ and $T_{\CAR}$. The data are
the same as in Fig.~\ref{fig:des_Sab}. The different points correspond to
different lengths $N$ along $Ox$ axis [see $N$ in
  Fig.~\ref{fig:schema3}(b)]. The blue dashed lines show the linear fits
$dS_{a,b}/dV=8T_{\EC}$, and $dS_{a,b}/dV=50 T_{\CAR}$.
\label{fig:des_Sab_vs_TEC_TCAR}
}
\end{figure*}

Considering the numerical data, plots of $d S_{a,b}/dV$ as functions of
$T_{\EC}$ and $T_{\CAR}$ (see Fig.~\ref{fig:des_Sab_vs_TEC_TCAR}), and a
comparison between $T_{\ARARbar}^{A,A}$ and $T_{\EC}$ confirm
Eqs.~\eqref{eq:ID1} and~\eqref{eq:ID2} and show in addition 
\begin{equation}
\frac{d S_{a,b}}{dV}\simeq 50 \frac{e^3}{h} T_{\CAR}\label{eq:id2}
.
\end{equation}
Comparing with the previous BTK calculation, it is suggested that
Eq.~\eqref{eq:id2} is model-dependent, because $T_{\CAR}=0$ for the BTK model
while $T_{\CAR}$ is finite but small in the self-consistent microscopic
calculation. On the other hand, Eq.~\eqref{eq:ID1}, which is obtained
also for the BTK model, is expected to be a fundamental relation. 

\begin{figure*}
  \centerline{\includegraphics[width=0.9\textwidth]{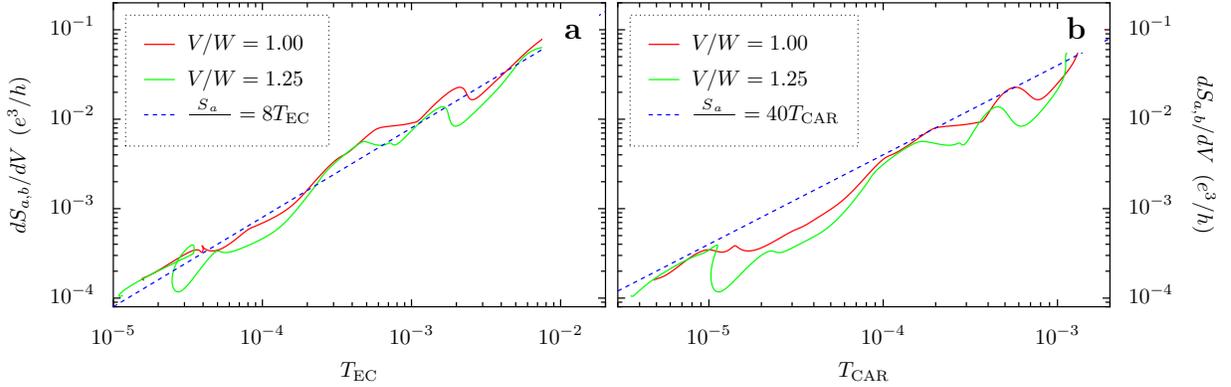}}
\caption{(Color online). 
Relation between $dS_{a,b}/dV$ and the transmission coefficients $T_{\EC}$ and
$T_{\CAR}$ in the presence of weak disorder ($V/W=1$ and $V/W=1.25$): Panel (a)
shows the relation between $dS_{a,b}/dV$ and $T_{\EC}$ and panel (b) shows the
relation between $dS_{a,b}/dV$ and $T_{\CAR}$. 
The figure is obtained by averaging over the contributions between the
combinations of all different transverse sites $x_n$ and $x_m$ in the two normal
leads.
The width of the model is given by $M=20$ [see $M$ in Fig.~\ref{fig:schema3}(b)].
\label{fig:des_Sab_vs_TEC_TCAR_des}
}
\end{figure*}

Until now, we only studied perfect systems without disorder. The numerical
calculations based on microscopic Green functions give us however the additional
possibility to add disorder to the model, which is an unavoidable ingredient to
describe experiments.

Eq.~\eqref{eq:ID1} remains approximately valid in the presence of weak
disorder (see Fig.~\ref{fig:des_Sab_vs_TEC_TCAR_des}), introduced in the form of
a random on-site potential uniformly
distributed in the interval $\left[-V,V\right]$, with elastic mean free path
$l_e \sim (W/V)^2 a_0$ smaller than the ballistic coherence length $\sim
(W/\Delta)a_0$, but not small compared to the Fermi wave-length.
The parameter $W$ is the hopping amplitude in the bulk of the
superconductor, taking the same value at the interfaces because of highly
transparent contacts. 

The coherence length is reduced as the strength of disorder increases. The
coherence lengths are fitted to $\xi=3.8a_0,\,3.6 a_0$ for $V/W=1.00,\,1.25$,
as compared to $\xi=8 a_0$ with $V/W=0$ in the ballistic limit. The coherence
length in the presence of disorder becomes smaller than its ballistic value,
as for a superconductor in the dirty limit. Within error-bars, Eq.~\eqref{eq:ID1}
is fulfilled also in the presence of weak disorder, while the coefficient in
Eq.~\eqref{eq:ID2} is changed, resulting in 
\begin{equation}
\frac{d S_{a,b}}{dV}\simeq 40 \frac{e^3}{h} T_{\CAR}.
\end{equation}
It is concluded that $T_{\CAR}\ll T_{\EC}$ implies that $S_{a,b}\simeq
S_{a,b}^{\ARARbar}$ and that Eq.~\eqref{eq:ID1} is fulfilled. Thus, it was shown
that, in this parameter regime, $d S_{a,b}/dV>0$ is evidence for {\ARARbar}, not
for {\CAR}.

The BTK approach and the
microscopic calculations lead to positive $d S_{a,b}/dV>0$, which is due to
the exchange of fermionic quasi-particles (see Fig.~\ref{fig:GpmGmp}). This is
the main physical result of our article: $dS_{a,b}/dV>0$ at high transparency is
not due to CAR. It is due to pairs of electron-like quasi-particles, pairs of
hole-like quasi-particles, and exchange of fermions.

\section{Conclusions}
\label{sec:conclu}

The article was already summarized in Sec.~\ref{sec:open-questions} and thus
we conclude with a brief overview and final remarks. We have evaluated
current-current cross-correlations in a NSN structure with a homogeneous
superconductor (without self-consistency in the order parameter), and with
strong inverse proximity effect (with self-consistent microscopic calculations
for a two-dimensional three-terminal set-up). For both approaches, the linear
differential cross-correlations $dS_{a,b}/dV$ are positive for highly
transparent contacts and decay exponentially with a characteristic length set
by the coherence length. Positive $dS_{a,b}/dV$ arises in this set-up not because of {\CAR},
but because of what is identified as the correlated penetration of pairs of
electron-like quasi-particles and pairs of hole-like quasi-particles into the superconductor in the form of
{\ARARbar}. The positive sign of $d S_{a,b}/dV$ is due to the additional exchange of two
fermions. It is emphasized that the proposed mechanism does not involve
quartets in the superconductor, because {\ARAR} does not contribute to the
current-current cross-correlations. Direct evaluation of $dS_{a,b}/dV$ leads
to $dS_{a,b}/dV=4(e^3/h) T^{A,A}_{\ARARbar}$, and to
$T^{A,A}_{\ARARbar}\simeq T_{EC}$, which holds also for a superconductor
with weak disorder and with elastic mean free path shorter than the coherence
length.

Finally, correlations between pairs of Andreev pairs were discussed
\cite{Nazarov,Heikkila} in connection with noise in an Andreev
interferometer. 

Future studies of this class of set-ups might include the effect of arbitrary
applied voltages and quantitative effects of three-dimensional geometry.

\section*{Acknowledgments}
The authors acknowledge fruitful discussions with B. Dou\c{c}ot, D. Feinberg,
M. Houzet, F. Lefloch, P. Samuelsson, R. Whitney. R. M. is grateful to
S. Bergeret and A. Levy Yeyati for a previous collaboration,\cite{Melin-Madrid}
and thanks the latter for useful remarks on a preliminary version of the
manuscript. We thank one of the referees for pointing out to us the possibility
to perform the integrals within the BTK model analytically.
Support from contract ANR P-NANO ELEC-EPR is acknowledged.

\appendix
\section{Details on the scattering approach}
\label{app:details-scat}
The elements $s_{i,j}^{\alpha,\beta}$ of the scattering matrix are calculated
within the BTK approach\cite{BTK} using two-component wave-functions describing
electrons and holes respectively. The indices $i,j$ refer to the normal
electrodes N$_a$ and N$_b$, while $\alpha,\beta$ run over the two components
describing the electrons and holes.

For example, the wave-functions for an electron incoming from electrode N$_a$
take the form
\begin{align}
\displaybreak[0]
\psi_a(x)&=
\begin{aligned}[t]
&\begin{pmatrix}1\\0\end{pmatrix}\left( e^{iq^{(+)}x}
+s^{e,e}_{a,a} e^{-iq^{(+)}x}\right)\\
&+s^{h,e}_{a,a} \begin{pmatrix}0\\1\end{pmatrix} e^{iq^{(-)}x}\,,
\end{aligned}\\
\displaybreak[0]
\psi_S(x)&=
\begin{aligned}[t]
&\begin{pmatrix}u_0\\v_0 \end{pmatrix} \left( c_1 e^{ik^{(+)}x}
+c'_1 e^{-ik^{(+)}(x-R)}\right)\\
&+\begin{pmatrix}v_0\\u_0 \end{pmatrix} \left( d_1 e^{-ik^{(-)}x}
+d'_1 e^{ik^{(-)}(x-R)}\right)\,,
\end{aligned}\\
\psi_b(x)&=
\begin{aligned}[t]
&\begin{pmatrix}0\\1\end{pmatrix} s^{h,e}_{b,a} e^{-iq^{(-)}(x-R)}
+\begin{pmatrix}1\\0\end{pmatrix} s^{e,e}_{b,a} e^{iq^{(+)}(x-R)}
,
\end{aligned}
\end{align}
where $\psi_a(x)$, $\psi_S(x)$ and $\psi_b(x)$ are the parts of the
wave-functions in the electrodes N$_a$, S, and N$_b$ respectively [see
Fig.~\ref{fig:schema3}(a)]. The notations $q^{(+)}$, $q^{(-)}$, $k^{(+)}$ and
$k^{(-)}$ stand for the wave-vectors in the normal and superconducting
electrodes:
\begin{align}
k^{(\pm)}&= k_F\pm i /\xi \\
q^{(\pm)}&= k_F\,,
\end{align}
with $k_F \xi \gg 1$.

The elements $s^{e,e}_{a,a}$, $s^{h,e}_{a,a}$, $s^{h,e}_{b,a}$ and
$s^{e,e}_{b,a}$ can be determined using the continuity of the wave-functions at
the interfaces
[$\psi_a(0)=\psi_S(0)$ and $\psi_S(R)=\psi_b(R)$] and the boundary condition for
the derivatives [$\psi_S'(0)-\psi_a'(0) = H \psi_a(0)$ and
$\psi_b'(R)-\psi_S'(R) = H \psi_b(R)$].
The BTK parameter $Z$ is defined by $Z=H/\hbar v_F$.

The remaining elements of the scattering matrix can be obtained from the other
possible scattering processes (e.g., a hole incoming from electrode N$_b$) by
analogous calculation.

By comparing the different equations, the symmetry 
\begin{equation}
s_{a,a}^{e,h}(\omega)=\overline{s_{b,b}^{h,e}(-\omega)}
\end{equation}
of the scattering matrix can be obtained.

The BCS coherence factors $u_0$, $v_0$ appearing in these equations are given by:
\begin{equation}
\label{eq:u0-v0}
u_0^2=1-v_0^2=\frac{1}{2}\left(1+\frac{\sqrt{\omega^2-\Delta^2}}
{\omega}\right)
.
\end{equation}
The coherence factors $u_0$ and $v_0$ are interchanged under complex
conjugation and changing sign of the real part of energy (a small
imaginary part is supposed to be added to $\omega$).

\section{Current-current cross-correlations for the BTK model in the limit
\boldmath $Z\simeq 0$}
\label{app:FaaFab2e}
For highly transparent interfaces ($Z\simeq 0$), 
the expressions obtained for
$s_{i,j}^{\alpha,\beta}$ simplify considerably, local reflections 
($s_{i,i}^{\alpha,\alpha}=0$) and non-local Andreev reflections
($s_{i,j}^{\alpha,\beta} =0$, with $i\neq j, \alpha \neq \beta$) are suppressed.
\cite{Zaikin}
Thus, only local Andreev reflection ($s_{i,i}^{\alpha,\beta}$, with
$\alpha\neq\beta$) and transmission without branch-crossing
($s_{i,j}^{\alpha,\alpha}$, with $i\neq j$) can occur. The non-zero elements of
the scattering matrix are given by
\begin{align}
  s_{a,a}^{h,e} &= u_0 v_0 \frac{e^{ R/\xi}-e^{- R/\xi}}
	{v_0^2 e^{R/\xi}-u_0^2e^{-R/\xi}}\,,\\
  s_{b,a}^{e,e} &= e^{-i k_F R} \frac{u_0^2 - v_0^2} {u_0^2 e^{-R/\xi} - v_0^2
e^{R/\xi}}\,.
\end{align}

Assuming $s(\omega)$ to be constant in the range $[-eV,eV]$ (we study the case
$eV\ll\Delta$), Eqs.~\eqref{eq:couranti} and~\eqref{eq:Sij} can be written as 
\begin{align}
\label{eq:courant-s-gene}
I_a&=\frac{4e^2}{h} V \left|s^{e,h}_{a,a}\right|^2\,,\\
\label{eq:noise-s-gene}
S_{a,b}&=-\frac{4e^3}{h}|V| \left(
s_{a,b}^{h,h\dagger}s_{b,b}^{h,e}s_{b,a}^{e,e}s_{a,a}^{e,h\dagger}
+
s_{a,a}^{h,e\dagger}s_{a,b}^{e,e}s^{e,h}_{b,b}s^{h,h\dagger}_{b,a}
\right)
\end{align}
for $i\ne j$.
In the limit $R/\xi\ll1$, the elements of the scattering matrix (evaluated for
$\omega\rightarrow 0$) are given by
\begin{align}
  s_{a,a}^{h,e} &= s_{b,b}^{e,h}\simeq i \frac{R}{\xi}\,,\\
  s_{b,a}^{e,e} s_{a,b}^{h,h}&\simeq 1,
\end{align}
leading to 
\begin{equation}
\label{eq:Saabb}
S_{a,a}=S_{a,b}=\frac{8 e^3}{h}|V|\left(\frac{R}{\xi}\right)^2 \geq
0\,.
\end{equation}
The result for $S_{a,a}$ is obtained by an analogous calculation.

Combining Eq.~\eqref{eq:courant-s-gene} for $I_a$ with Eq.~\eqref{eq:Saabb}
for $S_{a,a}$ and $S_{b,b}$ leads to $F_{a,a}\simeq F_{b,b}\simeq 1$ for
$eV\ll \Delta_0$,  $T_N=1$ and $R/\xi\ll 1$. 

\section{Technical details on microscopic calculations}
\label{app:micro-details}
\begin{figure*}
\centerline{\includegraphics[width=0.9\textwidth]{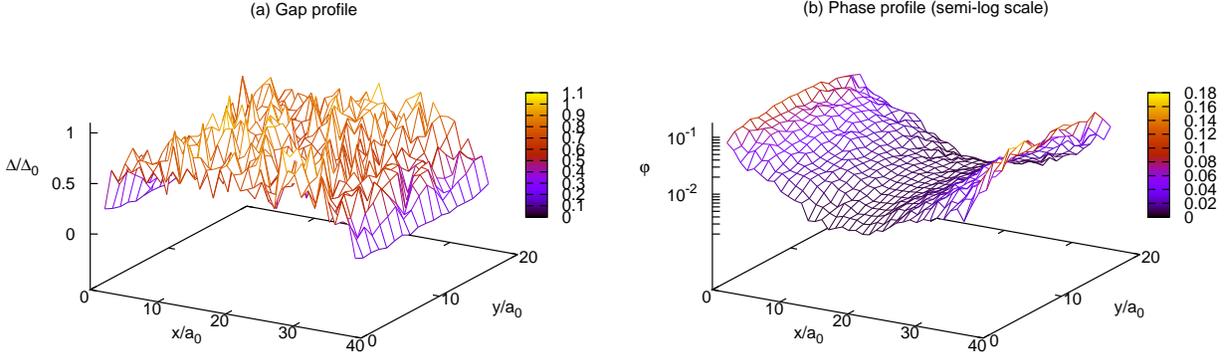}}
\caption{(Color online). The figure shows the gap profile in panel (a), the
  phase profile in semi-log scale in panel (b), for a given realization of
  disorder with $V/W=1$. See the text for the meaning of the parameters $V$ and
  $W$. The $x$ and $y$ axis are the same as in Fig.~\ref{fig:schema3}(b). This
  figure is obtained with the self-consistent algorithm developed in
  Ref.~\onlinecite{Melin-Madrid} for the ballistic case. The phase in panel (b)
  is small but non-zero because the algorithm takes into account
  non-equilibrium effects.
 \label{fig:des_gap}
}
\end{figure*}
The microscopic calculations are based on the following tight-binding
Hamiltonian on a square lattice:
\begin{equation}
\label{eq:H-tight-binding}
\begin{aligned}
{\cal H}_S=&-W\sum_{\langle n,m \rangle} \sum_\sigma
\left(c_{n,\sigma}^+ c_{m,\sigma}+
c_{m,\sigma}^+ c_{n,\sigma}\right)\\
&+ \sum_n \Delta_n
\left(c_{n,\uparrow}^+ c_{n,\downarrow}^+
+c_{n,\downarrow} c_{n,\uparrow}\right)\\
&-\sum_{n,\sigma} V_n c_{n,\sigma}^+ c_{n,\sigma}
,
\end{aligned}
\end{equation}
with the hopping amplitude $W$ between nearest neighbor sites $n$ and $m$
separated by a distance $a_0$. The normal electrodes are described by an
analogous
Hamiltonian with no pairing term and no disorder. Highly transparent contacts
with interfacial hopping equal to $W$ are used in Sec.~\ref{sec:micro}. The
parameter $\Delta_n$ is the superconducting order parameter at site $n$. It is
determined self-consistently in Sec.~\ref{sec:micro} on the basis of the
recursive algorithm developed in Ref.~\onlinecite{Melin-Madrid}. The gap in the
superconducting reservoirs takes the fixed value $\Delta_0$. Disorder is
introduced at the end of Sec.~\ref{sec:micro} in the form of a random on-site
potential $V_n$ on each tight-binding site, uniformly distributed in the
interval $\left[-V,V\right]$. The gap and phase profiles in the
superconducting island are shown in Fig.~\ref{fig:des_gap} in order to
illustrate the output of the part of the code performing the self-consistent
calculation. Because of disorder, the gap fluctuates strongly from one
tight-binding site to the next. The phase profile shows a smooth exponential
decay from the NS interfaces. The accuracy of the self-consistent calculation
gives access to variations of the phase over almost two orders of magnitude.

The average current in Eq.~\eqref{eq:Ia} is evaluated from the Keldysh Green
function:\cite{Cuevas}
\begin{equation}
\label{eq:I-a-alpha}
\begin{aligned}
I_{a,\alpha}=\frac{2e}{h}\sum_m \int d\omega \Big[& \hat{t}_{a_m,\alpha_m}
  \hat{G}^{+,-}_{\alpha_m,a_m}(\omega)\\
&- \hat{t}_{\alpha_m,a_m} \hat{G}^{+,-}_{a_m,\alpha_m}(\omega)\Big]_{1,1}
.
\end{aligned}
\end{equation}

Zero frequency noise (see Sec.~\ref{sec:hamil}) is obtained from
$S_{a,a}=2e^2t^2/h \int d\omega S_{a,a}(\omega)$ and $S_{a,b}=2e^2t^2/h \int d\omega
S_{a,b}(\omega)$, with
\begin{align}
\label{eq:Saa}
S_{a,a}(\omega)=&
\sum_{n,m}
\mbox{Tr}\Big[
\begin{aligned}[t]
&\hat{G}^{+,-}_{a_m,a_n}(\omega) \hat{G}^{-,+}_{\alpha_n,\alpha_m}(\omega)\\
&+\hat{G}^{+,-}_{\alpha_m,\alpha_n}(\omega) \hat{G}^{-,+}_{a_n,a_m}(\omega)\\
&-\hat{G}^{+,-}_{a_m,\alpha_n}(\omega) \hat{G}^{-,+}_{a_n,\alpha_m}(\omega)\\
&-\hat{G}^{+,-}_{\alpha_m,a_n}(\omega) \hat{G}^{-,+}_{\alpha_n,a_m}(\omega)
\Big]\,,
\end{aligned}
\\
\label{eq:Sab}
S_{a,b}(\omega)=&
\sum_{n,m}
\mbox{Tr}\Big[ 
\begin{aligned}[t]
&\hat{G}^{+,-}_{b_m,a_n}(\omega) \hat{G}^{-,+}_{\alpha_n,\beta_m}(\omega)\\
&+\hat{G}^{+,-}_{\beta_m,\alpha_n}(\omega) \hat{G}^{-,+}_{a_n,b_m}(\omega)\\
&-\hat{G}^{+,-}_{b_m,\alpha_n}(\omega) \hat{G}^{-,+}_{a_n,\beta_m}(\omega)\\
&-\hat{G}^{+,-}_{\beta_m,a_n}(\omega) \hat{G}^{-,+}_{\alpha_n,b_m}(\omega)
\Big]
.
\end{aligned}
\end{align}
The trace is evaluated over the Nambu labels. Eq.~\eqref{eq:Sab} is a
generalization of Ref.~\onlinecite{Cuevas-noise} to two interfaces with many
channels. The numerical calculations presented in the main body of the article
are based on Eqs.~\eqref{eq:I-a-alpha}, \eqref{eq:Saa} and~\eqref{eq:Sab}.

\begin{widetext}
\section{Microscopic calculations for the noise formula}
\label{app:micro-noise}
In this Appendix we provide the complete formula for the noise $S_{a,b}(\omega)$
given by Eq.~\eqref{eq:Sab} for sub-gap voltage 
($|\omega|<eV<\Delta$). This allows considerable simplification of the
expression of $S_{a,b}$ because the Keldysh Green function $g^{+-}=0$ in 
the isolated superconductor, as there exist no single-electron states.  The
total noise can be decomposed into different terms according to the types of
transmission modes in the superconductor as:
\begin{equation}
S_{a,b}(\omega)=S_{\CAR}+S_{\EC}+S_{\ARARbar}
+ S_{\ARAR}+S_\mathrm{MIXED}+S'
.
\end{equation}
In the normal lead $N$, the Keldysh Green functions read
$g^{+-,11}_{NN}=2i\pi\rho_Nn_F(\omega-eV)$,
$g^{+-,22}_{NN}=2i\pi\rho_Nn_F(\omega+eV)$,
$g^{-+,11}_{NN}=2i\pi\rho_Nn_F(-\omega+eV)$ and
$g^{-+,22}_{NN}=2i\pi\rho_Nn_F(-\omega-eV)$, where $\rho_N$ is the density of
states at the interface in the normal lead, and, at zero temperature,
$n_F(x)=\theta(-x)$, with $\theta(x)$ being the Heaviside step-function. This
gives $g^{+-,11}_{N,N}g^{-+,22}_{N,N}=g^{+-,22}_{N,N}g^{-+,11}_{N,N}=0$, leading to further
simplification of the expression for $S_{a,b}$. 

The contribution
$S_{\CAR}$ is given by the advanced-advanced or retarded-retarded transmission
modes in the electron-hole channel, in the form of the combinations of the
type $G^{A,12}_{\alpha\beta} G^{R,21}_{\beta\alpha}$. Physically, these
microscopic processes can be interpreted as Cooper pair splitting as appearing
in {\CAR}. The expression for $S_{\CAR}$ is as follows:
\begin{equation}
  \begin{aligned}
  S_{\CAR} = 2 t^4 \Big[
  &+g^{-+,11}_{bb} g^{+-,22}_{aa} G^{R,21}_{\alpha\beta} G^{R,12}_{\beta\alpha} 
  	(1 + 2 i G^{A,22}_{\alpha\alpha} \pi t^2 \rho_a) 
	(1 + 2 i G^{A,11}_{\beta\beta} \pi t^2 \rho_b)\\
  &+g^{-+,22}_{bb} g^{+-,11}_{aa} G^{R,12}_{\alpha\beta} G^{R,21}_{\beta\alpha}
  	(1 + 2 i G^{A,11}_{\alpha\alpha} \pi t^2 \rho_a) 
	(1 + 2 i G^{A,22}_{\beta\beta} \pi t^2 \rho_b)\\
  &-g^{-+,22}_{aa} g^{+-,11}_{bb} G^{A,21}_{\alpha\beta} G^{A,12}_{\beta\alpha}  
  	(i + 2 G^{R,22}_{\alpha\alpha} \pi t^2 \rho_a) 
	(i + 2 G^{R,11}_{\beta\beta} \pi t^2 \rho_b)\\
  &-g^{-+,11}_{aa} g^{+-,22}_{bb} G^{A,12}_{\alpha\beta} G^{A,21}_{\beta\alpha} 
  	(i + 2 G^{R,11}_{\alpha\alpha} \pi t^2 \rho_a)
	(i + 2 G^{R,22}_{\beta\beta} \pi t^2 \rho_b)
  \Big]\,,
  \end{aligned}
\end{equation}
where $t$ is the hopping amplitude at the interfaces.

The contribution $S_{\EC}$ contains advanced-advanced and
retarded-retarded transmission modes in the electron-electron or hole-hole
channel, in the form of the combinations of the type $G^{A,12}_{\alpha\beta}
G^{A,21}_{\beta\alpha}$. This is the contribution to the noise of normal
electron transmission in the form of {\EC}. The expression reads
\begin{equation}
  \begin{aligned}
  S_{\EC} = 2 t^4 \Big[
  &+g^{-+,11}_{bb} g^{+-,11}_{aa} G^{R,11}_{\alpha\beta} G^{R,11}_{\beta\alpha} 
  	(i - 2 G^{A,11}_{\alpha\alpha} \pi t^2 \rho_a) 
	(i - 2 G^{A,11}_{\beta\beta} \pi t^2 \rho_b)\\
  &+g^{-+,22}_{bb} g^{+-,22}_{aa} G^{R,22}_{\alpha\beta} G^{R,22}_{\beta\alpha} 
   	(i - 2 G^{A,22}_{\alpha\alpha} \pi t^2 \rho_a) 
	(i - 2 G^{A,22}_{\beta\beta} \pi t^2 \rho_b)\\
  &+g^{-+,11}_{aa} g^{+-,11}_{bb} G^{A,11}_{\alpha\beta} G^{A,11}_{\beta\alpha} 
   	(i + 2 G^{R,11}_{\alpha\alpha} \pi t^2 \rho_a) 
	(i + 2 G^{R,11}_{\beta\beta} \pi t^2 \rho_b)\\
  &+g^{-+,22}_{aa} g^{+-,22}_{bb} G^{A,22}_{\alpha\beta} G^{A,22}_{\beta\alpha}  
	(i + 2 G^{R,22}_{\alpha\alpha} \pi t^2 \rho_a)
	(i + 2 G^{R,22}_{\beta\beta} \pi t^2 \rho_b)
  \Big]\,.
  \end{aligned}
\end{equation}

Now we consider processes that do not appear in lowest order perturbation
theory. First the contribution $S_{\ARARbar}$ contains advanced-advanced and
retarded-retarded transmission modes, in the form of the combinations of the
type $G^{A,11}_{\alpha\beta} G^{A,22}_{\beta\alpha}$. Physically, this can be
interpreted as the contribution to the noise of synchronized Andreev and
inverse Andreev processes ({\ARARbar}), with the exchange of two fermions at the
same time, thus leading to $dS_{a,b}/dV>0$ (see Fig.~\ref{fig:GpmGmp}). This
contribution dominates the current-current cross-correlations $dS_{a,b}/dV$ in
the considered set-up. This contribution reads
\begin{equation}
\label{eq:SARARbar}
  \begin{aligned}
  S_{\ARARbar} = -8 \pi^2 t^8 \rho_a \rho_b \Big(
  &+g^{-+,22}_{bb} g^{+-,11}_{aa} G^{A,12}_{\alpha\alpha} G^{A,21}_{\beta\beta} G^{R,22}_{\alpha\beta} G^{R,11}_{\beta\alpha}
   +g^{-+,11}_{bb} g^{+-,22}_{aa} G^{A,21}_{\alpha\alpha} G^{A,12}_{\beta\beta} G^{R,11}_{\alpha\beta} G^{R,22}_{\beta\alpha}\\
  &+g^{-+,11}_{aa} g^{+-,22}_{bb} G^{A,11}_{\alpha\beta} G^{A,22}_{\beta\alpha} G^{R,21}_{\alpha\alpha} G^{R,12}_{\beta\beta}
   +g^{-+,22}_{aa} g^{+-,11}_{bb} G^{A,22}_{\alpha\beta} G^{A,11}_{\beta\alpha} G^{R,12}_{\alpha\alpha} G^{R,21}_{\beta\beta}
  \Big)\,.
  \end{aligned}
\end{equation}

Another contribution not appearing in lowest order contains processes
involving electron-hole conversion both at the interfaces and during
propagation in the superconductor. The transmission modes in the
superconductor are of the type
$G^{A,12}_{\alpha,\beta}G^{A,12}_{\beta,\alpha}$. These terms correspond to
the synchronization of two Andreev processes ({\ARAR}). The corresponding
expression reads
\begin{equation}
  \begin{aligned}
  S_{\ARAR} = 8 \pi^2 t^8 \rho_a \rho_b \Big(
  &+g^{-+,22}_{bb} g^{+-,22}_{aa} G^{A,21}_{\alpha\alpha} G^{A,21}_{\beta\beta} G^{R,12}_{\alpha\beta} G^{R,12}_{\beta\alpha}
   +g^{-+,11}_{bb} g^{+-,11}_{aa} G^{A,12}_{\alpha\alpha} G^{A,12}_{\beta\beta} G^{R,21}_{\alpha\beta} G^{R,21}_{\beta\alpha}\\ 
  &+g^{-+,22}_{aa} g^{+-,22}_{bb} G^{A,21}_{\alpha\beta} G^{A,21}_{\beta\alpha} G^{R,12}_{\alpha\alpha} G^{R,12}_{\beta\beta} 
   +g^{-+,11}_{aa} g^{+-,11}_{bb} G^{A,12}_{\alpha\beta} G^{A,12}_{\beta\alpha} G^{R,21}_{\alpha\alpha} G^{R,21}_{\beta\beta}
  \Big)\,.
  \end{aligned}
\end{equation}

Another contribution contains transmission modes of the type
$G_{\alpha,\beta}^{A,11}G_{\beta,\alpha}^{A,12}$. These terms read
\begin{align}
\displaybreak[0]
  S' = & 8\pi^2 t^8 \rho_a \rho_b \Big[
  \begin{aligned}[t]
  &+g^{+-,11}_{aa} G^{A,11}_{\alpha\alpha} \left(g^{-+,22}_{bb} G^{A,21}_{\beta\beta} G^{R,12}_{\alpha\beta} G^{R,11}_{\beta\alpha}
						-g^{-+,11}_{bb} G^{A,12}_{\beta\beta} G^{R,11}_{\alpha\beta} G^{R,21}_{\beta\alpha}\right)\\ 
  &+g^{+-,11}_{aa} G^{A,12}_{\alpha\alpha} \left(g^{-+,22}_{bb} G^{A,22}_{\beta\beta} G^{R,22}_{\alpha\beta} G^{R,21}_{\beta\alpha}
						-g^{-+,11}_{bb} G^{A,11}_{\beta\beta} G^{R,21}_{\alpha\beta} G^{R,11}_{\beta\alpha}\right)\\ 
  &+g^{+-,22}_{aa} G^{A,21}_{\alpha\alpha} \left(g^{-+,11}_{bb} G^{A,11}_{\beta\beta} G^{R,11}_{\alpha\beta} G^{R,12}_{\beta\alpha}
						-g^{-+,22}_{bb} G^{A,22}_{\beta\beta} G^{R,12}_{\alpha\beta} G^{R,22}_{\beta\alpha}\right)\\
  &+g^{+-,22}_{aa} G^{A,22}_{\alpha\alpha} \left(g^{-+,11}_{bb} G^{A,12}_{\beta\beta} G^{R,21}_{\alpha\beta} G^{R,22}_{\beta\alpha}
						-g^{-+,22}_{bb} G^{A,21}_{\beta\beta} G^{R,22}_{\alpha\beta} G^{R,12}_{\beta\alpha}\right)\\
  &+g^{-+,22}_{aa} G^{A,21}_{\alpha\beta} \left(g^{+-,11}_{bb} G^{A,11}_{\beta\alpha} G^{R,12}_{\alpha\alpha} G^{R,11}_{\beta\beta} 
					       -g^{+-,22}_{bb} G^{A,22}_{\beta\alpha} G^{R,22}_{\alpha\alpha} G^{R,12}_{\beta\beta}\right)\\ 
  &+g^{-+,22}_{aa} G^{A,22}_{\alpha\beta} \left(g^{+-,11}_{bb} G^{A,12}_{\beta\alpha} G^{R,22}_{\alpha\alpha} G^{R,21}_{\beta\beta} 
					       -g^{+-,22}_{bb} G^{A,21}_{\beta\alpha} G^{R,12}_{\alpha\alpha} G^{R,22}_{\beta\beta}\right)\\ 
  &+g^{-+,11}_{aa} G^{A,11}_{\alpha\beta} \left(g^{+-,22}_{bb} G^{A,21}_{\beta\alpha} G^{R,11}_{\alpha\alpha} G^{R,12}_{\beta\beta} 
					       -g^{+-,11}_{bb} G^{A,12}_{\beta\alpha} G^{R,21}_{\alpha\alpha} G^{R,11}_{\beta\beta}\right)\\ 
  &+g^{-+,11}_{aa} G^{A,12}_{\alpha\beta} \left(g^{+-,22}_{bb} G^{A,22}_{\beta\alpha} G^{R,21}_{\alpha\alpha} G^{R,22}_{\beta\beta}
					       -g^{+-,11}_{bb} G^{A,11}_{\beta\alpha} G^{R,11}_{\alpha\alpha} G^{R,21}_{\beta\beta}\right) 
  \Big]\end{aligned}\\
\displaybreak[0]
\nonumber
&+4 i \pi t^6 \rho_a \Big[
  \begin{aligned}[t]
  &+g^{-+,22}_{aa} G^{R,12}_{\alpha\alpha} \left(g^{+-,11}_{bb} G^{A,21}_{\alpha\beta} G^{A,11}_{\beta\alpha}  
						-g^{+-,22}_{bb} G^{A,22}_{\alpha\beta} G^{A,21}_{\beta\alpha}\right)\\ 
  &+g^{-+,11}_{aa} G^{R,21}_{\alpha\alpha} \left(g^{+-,22}_{bb} G^{A,12}_{\alpha\beta} G^{A,22}_{\beta\alpha}
						-g^{+-,11}_{bb} G^{A,11}_{\alpha\beta} G^{A,12}_{\beta\alpha}\right)\\
  &+g^{+-,11}_{aa} G^{A,12}_{\alpha\alpha} \left(g^{-+,11}_{bb} G^{R,21}_{\alpha\beta} G^{R,11}_{\beta\alpha} 
						-g^{-+,22}_{bb} G^{R,22}_{\alpha\beta} G^{R,21}_{\beta\alpha}\right)\\
  &+g^{+-,22}_{aa} G^{A,21}_{\alpha\alpha} \left(g^{-+,22}_{bb} G^{R,12}_{\alpha\beta} G^{R,22}_{\beta\alpha}
						-g^{-+,11}_{bb} G^{R,11}_{\alpha\beta} G^{R,12}_{\beta\alpha}\right)
  \Big]\end{aligned}\\
\nonumber
&+4 i \pi t^6 \rho_b \Big[
  \begin{aligned}[t]
  &+g^{-+,22}_{bb} G^{A,21}_{\beta\beta} \left(g^{+-,22}_{aa} G^{R,22}_{\alpha\beta} G^{R,12}_{\beta\alpha}
					      -g^{+-,11}_{aa} G^{R,12}_{\alpha\beta} G^{R,11}_{\beta\alpha}\right)\\
  &+g^{-+,11}_{bb} G^{A,12}_{\beta\beta} \left(g^{+-,11}_{aa} G^{R,11}_{\alpha\beta} G^{R,21}_{\beta\alpha}
					      -g^{+-,22}_{aa} G^{R,21}_{\alpha\beta} G^{R,22}_{\beta\alpha}\right)\\
  &+g^{+-,22}_{bb} G^{R,12}_{\beta\beta} \left(g^{-+,11}_{aa} G^{A,11}_{\alpha\beta} G^{A,21}_{\beta\alpha}
					      -g^{-+,22}_{aa} G^{A,21}_{\alpha\beta} G^{A,22}_{\beta\alpha}\right)\\
  &+g^{+-,11}_{bb} G^{R,21}_{\beta\beta} \left(g^{-+,22}_{aa} G^{A,22}_{\alpha\beta} G^{A,12}_{\beta\alpha}
					      -g^{-+,11}_{aa} G^{A,12}_{\alpha\beta} G^{A,11}_{\beta\alpha}\right)
  \Big]\,.\end{aligned}
\end{align}

The last term involves advanced-retarded transmission modes
\begin{equation}
  \begin{aligned}
  S_\mathrm{MIXED} =& 8 \pi^2 t^8 \rho_a \rho_b \Big[
  \begin{aligned}[t]
  &+g^{-+,22}_{aa} g^{+-,11}_{aa} 
	(G^{A,11}_{\alpha\alpha} G^{R,12}_{\alpha\alpha} - G^{A,12}_{\alpha\alpha} G^{R,22}_{\alpha\alpha}) 
	(G^{A,21}_{\alpha\beta} G^{R,11}_{\beta\alpha} - G^{A,22}_{\alpha\beta} G^{R,21}_{\beta\alpha})\\ 
  &+g^{-+,11}_{aa} g^{+-,22}_{aa} 
	(G^{A,21}_{\alpha\alpha} G^{R,11}_{\alpha\alpha} - G^{A,22}_{\alpha\alpha} G^{R,21}_{\alpha\alpha}) 
	(G^{A,11}_{\alpha\beta} G^{R,12}_{\beta\alpha} - G^{A,12}_{\alpha\beta} G^{R,22}_{\beta\alpha})\\
  &+g^{-+,22}_{bb} g^{+-,11}_{bb} 
	(G^{A,11}_{\beta\alpha} G^{R,12}_{\alpha\beta} - G^{A,12}_{\beta\alpha} G^{R,22}_{\alpha\beta}) 
	(G^{A,21}_{\beta\beta} G^{R,11}_{\beta\beta} - G^{A,22}_{\beta\beta} G^{R,21}_{\beta\beta})\\
  &+g^{-+,11}_{bb} g^{+-,22}_{bb} 
	(G^{A,21}_{\beta\alpha} G^{R,11}_{\alpha\beta} - G^{A,22}_{\beta\alpha} G^{R,21}_{\alpha\beta}) 
	(G^{A,11}_{\beta\beta} G^{R,12}_{\beta\beta} - G^{A,12}_{\beta\beta} G^{R,22}_{\beta\beta})
  \Big]
  \end{aligned}\\
  &-4 i \pi t^6 \rho_a \Big[
  \begin{aligned}[t]
  &+g^{-+,11}_{bb} g^{+-,22}_{bb} (G^{A,12}_{\beta\beta} + G^{R,12}_{\beta\beta})
	(G^{A,21}_{\beta\alpha} G^{R,11}_{\alpha\beta} - G^{A,22}_{\beta\alpha} G^{R,21}_{\alpha\beta})\\
  &-g^{-+,22}_{bb} g^{+-,11}_{bb} (G^{A,21}_{\beta\beta} + G^{R,21}_{\beta\beta})
	(G^{A,11}_{\beta\alpha} G^{R,12}_{\alpha\beta} - G^{A,12}_{\beta\alpha} G^{R,22}_{\alpha\beta}) 
  \Big]
  \end{aligned}\\
  &-4 i  \pi t^6 \rho_b \Big[
  \begin{aligned}[t]
  &+g^{-+,22}_{aa} g^{+-,11}_{aa} (G^{A,12}_{\alpha\alpha} + G^{R,12}_{\alpha\alpha}) 
	(G^{A,21}_{\alpha\beta} G^{R,11}_{\beta\alpha} - G^{A,22}_{\alpha\beta} G^{R,21}_{\beta\alpha})\\
  &-g^{-+,11}_{aa} g^{+-,22}_{aa} (G^{A,21}_{\alpha\alpha} + G^{R,21}_{\alpha\alpha}) 
	(G^{A,11}_{\alpha\beta} G^{R,12}_{\beta\alpha} - G^{A,12}_{\alpha\beta} G^{R,22}_{\beta\alpha})
  \Big]\,.
  \end{aligned}
\end{aligned}
\end{equation}

\section{Evaluation of the Green function \boldmath$\hat{G}_{\alpha,\beta}$
  connecting two interfaces}
\label{app:Galpha-beta}

It will be shown that, at small energy compared to the gap, the
advanced-advanced {\ARARbar} transmission mode
$\langle G_{\alpha,\beta}^{A,1,1}G_{\beta,\alpha}^{A,2,2}\rangle_{av}$ can be
replaced by the opposite of the advanced-retarded {\EC} transmission mode
$-\langle G_{\alpha,\beta}^{A,1,1}G_{\beta,\alpha}^{R,1,1}\rangle_{av}$. The
Green function is expanded in powers of the exponential coefficient
$\exp(-R_x/\xi)$ appearing in $\hat{g}_{\alpha,\beta}$, giving
\cite{Melin-Feinberg-PRB}
\begin{align}
\hat{G}^A_{\alpha,\beta}&= \hat{M}^A_{\alpha,\alpha}\hat{  g}^A_{\alpha,\beta}
\hat{N}^A_{\beta,\beta} + {\cal O}\left((g^A_{\alpha,\beta})^3\right)\\
\hat{M}^A_{\alpha,\alpha}&=\left(\hat{I}-\hat{g}^A_{\alpha,\alpha}\hat{t}_{\alpha,a}\hat{g}^A_{a,a}\hat{t}_{a,\alpha}\right)^{-1}\\
\hat{N}^A_{\beta,\beta}&=\left(\hat{I}-\hat{t}_{\beta,b}\hat{g}^A_{b,b}\hat{t}_{b,\beta}\hat{g}^A_{\beta,\beta}\right)^{-1}
.
\end{align}
This expansion leads to
\begin{equation}
\hat{G}^A_{\alpha,\beta}=\frac{1}{4}\left(\begin{array}{cc}
g_{\alpha,\beta}^{A,1,1}-g^{A,2,2}_{\alpha,\beta}
+i\left(g_{\alpha,\beta}^{A,1,2}+g^{A,2,1}_{\alpha,\beta}\right)
& i(g_{\alpha,\beta}^{A,1,1}+g_{\alpha,\beta}^{A,2,2})
+g_{\alpha,\beta}^{A,1,2}-g^{A,2,1}_{\alpha,\beta}
\\
i\left(g_{\alpha,\beta}^{A,1,1}+g_{\alpha,\beta}^{A,2,2}\right)
+g_{\alpha,\beta}^{A,2,1}-g^{A,1,2}_{\alpha,\beta}
& -g_{\alpha,\beta}^{A,1,1}+g_{\alpha,\beta}^{A,2,2}
+i\left(g_{\alpha,\beta}^{A,1,2}+g^{A,2,1}_{\alpha,\beta}\right)
\end{array}\right)
.
\end{equation}
One has $\hat g^A_{\alpha,\beta}=\hat g^R_{\alpha,\beta}\equiv \hat g_{\alpha,\beta}$
for energies within the gap.
The off-diagonal Nambu components are vanishingly small if $\omega\ll \Delta$.
It is deduced that
\begin{align}
\label{eq:A}
\langle G^{A,1,1}_{\alpha,\beta}G^{A,2,2}_{\beta,\alpha}\rangle_{av} &=
-\langle\left(g_{\alpha,\beta}^{1,1}-g_{\alpha,\beta}^{2,2}\right)^2\rangle_{av}
-\langle\left(g_{\alpha,\beta}^{1,2}+g_{\alpha,\beta}^{2,1}\right)^2\rangle_{av}\\
\langle G^{A,1,1}_{\alpha,\beta}G^{R,1,1}_{\beta,\alpha}\rangle_{av} &=
\langle
\left(g_{\alpha,\beta}^{1,1}-g_{\alpha,\beta}^{2,2}\right)^2\rangle_{av}
+\langle\left(g_{\alpha,\beta}^{1,2}+g_{\alpha,\beta}^{2,1}\right)^2\rangle_{av}
,
\end{align}
leading to the identification of the advanced-advanced {\ARARbar} transmission
coefficient to the opposite of the advanced-retarded {\EC} transmission
coefficient, for small energy compared to the gap and for $Na_0\agt
\xi$. Eq.~\eqref{eq:A} has no imaginary part because
$\langle g_{\alpha,\beta}^{1,2} g_{\alpha,\beta}^{2,2}\rangle_{av}=0$ at small
energy compared to the gap.

The fully dressed local Green function can be evaluated approximately by
inverting the Dyson equation
$\hat{G}_{\alpha,\alpha}=\hat{g}_{\alpha,\alpha}+\hat{g}_{\alpha,\alpha}\hat{t}_{\alpha,a}\hat{g}_{a,a}\hat{t}_{a,\alpha}\hat{G}_{\alpha,\alpha}$,
leading to $G_{\alpha,\alpha}^{A,1,2}=G_{\alpha,\alpha}^{A,2,1}=1/2W$ at
energy small compared to the gap.
\end{widetext}

\end{document}